\documentclass[twocolumn,10pt]{article}
\usepackage{geometry}
\usepackage{bm}
\usepackage{amsmath,amssymb,amsfonts}
\usepackage{graphicx}
\usepackage{textcomp}
\usepackage{multirow}
\usepackage{makecell}
\usepackage{etoolbox}
\usepackage[]{pifont}
\usepackage[noend]{algpseudocode}
\usepackage{algorithmicx,algorithm}
\usepackage[numbers,sort]{natbib}
\usepackage[labelfont=bf]{caption}
\usepackage{float}
\usepackage{multicol,lipsum}

\usepackage[T1]{fontenc}
\usepackage[utf8]{inputenc}
\usepackage{authblk}
\usepackage[colorlinks,linkcolor=blue,citecolor=blue,urlcolor=blue]{hyperref}
\usepackage[title]{appendix}
\usepackage{lettrine}
\usepackage{hyperref}

\graphicspath{{figures/}} 

\numberwithin{equation}{section}

\title{\textbf{{\LARGE Hierarchical Bayesian inference for community detection and connectivity of functional brain networks}}}

\title{\textbf{{\LARGE Hierarchical Bayesian inference for community detection and connectivity of functional brain networks}}}

\author[a,b,c,*]{\small Lingbin Bian }
\author[e]{\small Nizhuan Wang }
\author[c,d]{\small Leonardo Novelli }
\author[b]{\small Jonathan Keith }
\author[c,d,f,g]{\small Adeel Razi }

\affil[a]{School of Biomedical Engineering \& State Key Laboratory of Advanced Medical Materials and
Devices, ShanghaiTech University, China}
\affil[b]{School of Mathematics, Monash University, Australia}
\affil[c]{Turner Institute for Brain and Mental Health, School of Psychological Sciences, Monash University, Australia}
\affil[d]{Monash Biomedical Imaging, Monash University, Australia}
\affil[e]{Department of Chinese and Bilingual
Studies, The Hong Kong Polytechnic University, 
Hong Kong SAR, China}
\affil[f]{Wellcome Centre for Human Neuroimaging, University College London, United Kingdom}
\affil[g]{CIFAR Azrieli Global Scholars Program, CIFAR, Canada}

\affil[*]{\small Corresponding author: Lingbin Bian (lingbin.bian@gmail.com)}
\date{} 

\begin{document}
\renewcommand{\figurename}{\textbf{Fig.}} 
\maketitle



\begin{abstract}
\setlength{\parindent}{0pt} \setlength{\parskip}{1.5ex plus 0.5ex
minus 0.2ex} 
Most functional magnetic resonance imaging studies rely on estimates of hierarchically organized functional brain networks whose segregation and integration reflect the cognitive and behavioral changes in humans. However, most existing methods for estimating the community structure of networks from both individual and group-level analysis methods do not account for the variability between subjects. In this paper, we develop a new multilayer community detection method based on Bayesian latent block model (LBM). The method can robustly detect the community structure of weighted functional networks with an unknown number of communities at both individual and group levels and retain the variability of the individual networks. For validation, we propose a new community structure-based multivariate Gaussian generative model to simulate synthetic signal. Our simulation study shows that the community memberships estimated by hierarchical Bayesian inference are consistent with the predefined node labels in the generative model. The method is also tested via split-half reproducibility using working memory task fMRI data of 100 unrelated healthy subjects from the Human Connectome Project. Analyses using both synthetic and real data show that our proposed method is more accurate and reliable compared with the commonly used (multilayer) modularity models. The code of this work is available at: \underline{https://github.com/LingbinBian/CommuDetectLBM}.
\end{abstract}

\section{Introduction}
\lettrine[lines=3]{\textbf{C}}{ \textbf{ommunity}} 
detection techniques, with their ability to uncover hidden functional organization, have diverse applications, such as in neuroscience to analyze brain connectivity, as well as to study social network interactions on platforms such as Facebook or email communications \cite{Kawash2017}.

A community consists of nodes that are more densely linked to each other than to the rest of the network. There are mainly three key categories of community detection approaches across different fields. The first is clustering-based methods, such as spectral clustering \cite{Luxburg2007, Cribben2017}, which is a graph-based clustering technique that uses the eigenspectrum of the Laplacian matrix to perform dimensionality reduction before clustering in a lower-dimensional space. Spectral clustering typically requires predefining the number of communities or clusters similar to K-means clustering. The second category of community detection methods is based on maximization of modularity \cite{Newman2004, Newman2006} to find the best network partition such that there are more edges within the module than the expected number of edges by chance. For multiplex networks, multilyer modularity was proposed \cite{Mucha2010} to extend the concept of traditional modularity to systems with multiple layers or types of interactions, e.g. the brain networks of different subjects or the temporal dynamic brain networks. Both modularity and multilayer modularity have been widely used in computational neuroscience for exploring the complexity of brain organization such as in learning \cite{Bassett2011} and brain development \cite{Bian2025}. The third category is community detection based on the stochastic block model (SBM) \cite{Bian}. For example, One study integrated the SBM with non-overlapping sliding windows to estimate dynamic functional connectivity (FC) across networks. The edge weights were determined by averaging coherence matrices across subjects, and a threshold was used to convert the FC into a binary form \cite{Robinson2015}.

There are several problems with existing community detection studies. First, the number of communities is difficult to infer. For example, the spectral clustering method requires a predefined community number. Modularity and multilayer modularity have a parameter resolution to control the size and number of communities. However, how to fine-tune this resolution parameter and find the best one for the real world situation is still an open problem. Second, many studies may not retain complete information about signals using threshold strategies to binarize the network, for example the sliding windows based dynamic network study described above \cite{Robinson2015}. Third, most studies lack an evaluation of inter-subject variability. For example, the group-averaged adjacency matrix was calculated as an observation \cite{Bian2021}, which does not characterize the variability of FC between different subjects and also ignores the topological properties of the individual network of higher order.

To robustly detect the community structure of functional brain networks, recent studies have begun to focus on multiple networks from different subjects and estimate common features of network patterns using group-level analysis to reduce uncertainty caused by non-neural noise. For example, a method based on multi-subject stochastic block modelling can flexibly evaluate inter-individual variations in the community structure of functional networks \cite{Pavlovic2020}, but also treats the FC as binary edges. Other techniques that can capture brain networks at both the individual and group level by taking into account variations between subjects in BOLD time series using SBM or modularity include \cite{Ting2021} and \cite{Betzel2019}.

In this paper, we present a new multilayer community detection method based on hierarchical Bayesian inference to capture the variability between subjects at the group level. Here, a layer of community structure corresponds to an individual's functional brain network. We use the latent block model (LBM) to characterize individual-level FC and infer a latent label vector by Markov chain Monte Carlo (MCMC) sampling with an unknown number of communities with absorption and ejection strategy \cite{Nobile2007, Wyse2012} to estimate the subject-specific community structure that underlies a brain network architecture. At the group level, we model the estimated latent label vectors using a Categorical-Dirichlet conjugate pair and define a maximum label assignment probability matrix (MLAPM) that provides information about the group-level community structure. In addition, we model individual-level FC using a Normal-Normal-Inverse-Gamma (Normal-NIG) conjugate pair and estimate the mean and variance connectivity to characterize a group representative network. In Bayesian inference, the posterior distribution is obtained by combining prior and likelihood. If the prior and posterior follow from the same distribution family, then we can say that the prior is a conjugate for the likelihood or call them a prior-likelihood conjugate pair. 

An important issue in existing research is that, while much attention is given to community detection methods, the actual performance of detected communities is rarely verified. For that purpose, we simulate several sets of synthetic data using a multivariate Gaussian generative model, which also takes into account the inter-individual variation, where a covariance matrix encodes the ground truth of both group-level and subject-specific community labels and inter and intra community densities. The estimation of the community structure using our hierarchical Bayesian inference aligns well with the predefined community labels of both individual and group levels in the generative model used to simulate the synthetic data.

We further validate our method using working memory task fMRI data from the Human Connectome Project \cite{VanEssen2013, Barch2013} to estimate the community structures of brain network architectures and (weighted) connectivity. We used the split-half reproducibility strategy to evaluate the reliability of our method. The estimated community structures of brain network architectures show distinctive patterns between 2-back, 0-back, and fixation in a working memory task fMRI experiment.

In this paper, we are interested in the techniques that detect communities without predefining the number of communities. For that, we compared our LBM with the commonly used modularity and multilayer modularity model in both synthetic and working memory task fMRI data. Our approach achieves superior performance when measured by the normalized mutual information (NMI) between the detected communities and the ground truth compared with (multilayer) modularity models. 

This paper presents three main contributions: (i) We proposed a novel hierarchical Bayesian modelling method based on LBM and categorical-Dirichlet conjugate pair. Our method can estimate the community structure of individual- and group-level networks with an unknown number of communities while taking into account the inter-individual variability. (ii) We modelled the group-level brain connectivity based on Normal-Normal-Inverse-Gamma (Normal-NIG) conjugate pair that can capture both the mean strength and variance of group representative network. (iii) We proposed a novel generative model to simulate synthetic data with assumed group and subject-specific latent community labels and inter- and intra-community connectivity densities modelling the spatiotemporal segregation and integration of networks.

\section{Methodology}

\begin{figure*}[!ht]
\centering
\includegraphics[width=1\linewidth]{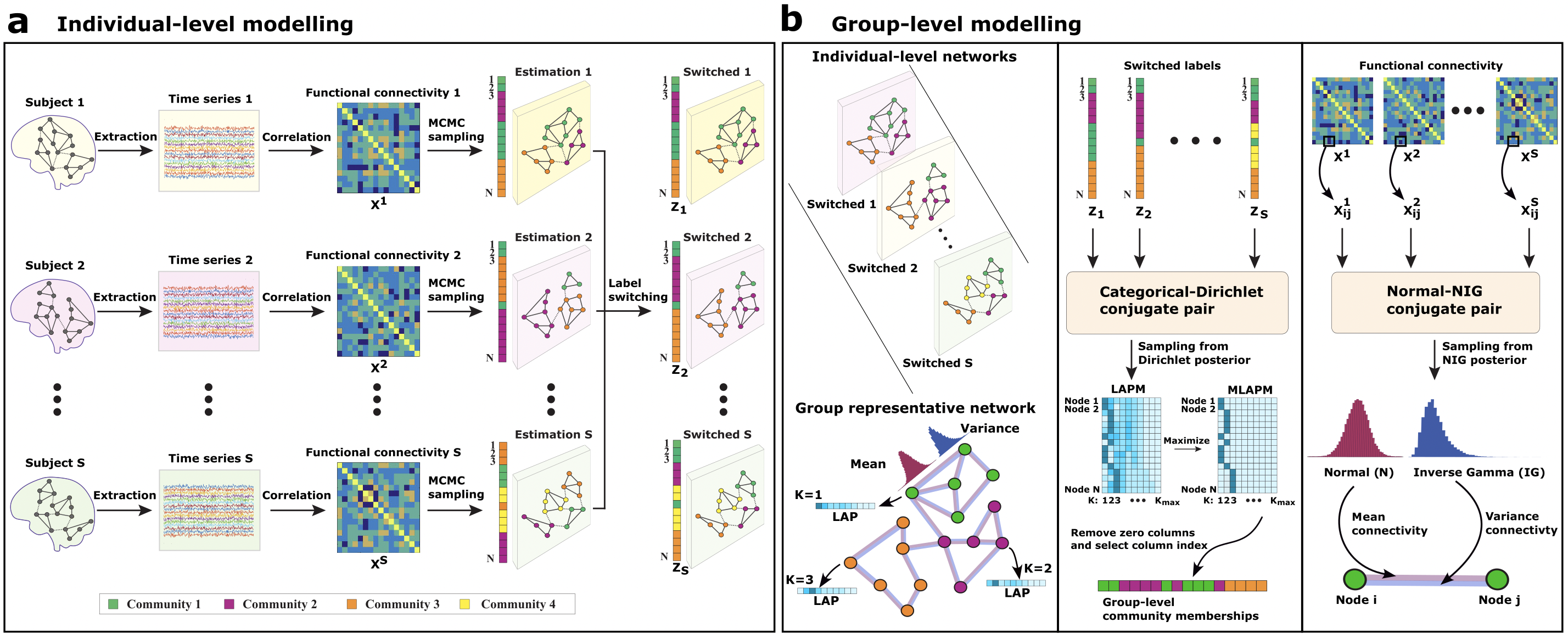}
\caption{\footnotesize Multilayer community detection for functional brain networks. \textbf{a} Individual-level modelling. For real data analysis, the BOLD time series are extracted from brain regions of interest for each subject. For synthetic data analysis, the time series are simulated from generative models with known community structures. The FC (a weighted adjacency matrix) is computed via Pearson's correlation from the regional time series. The community memberships of the network are estimated by drawing latent label vectors from a posterior density using MCMC sampling for each subject. A label switching method is applied to relabel the latent labels and make them consistent across subjects. \textbf{b} Group-level modelling. We analyze the group representative network from all of the subjects (from subject 1 to $S$). The community memberships $\lbrace\boldsymbol{z}_{1},\cdots,\boldsymbol{z}_{s}\rbrace$ estimated for each individual are treated as realizations of a latent group-level community structure. A Bayesian model with categorical-Dirichlet pair is used to fit the community memberships and a maximum label assignment probability matrix (MLAPM) is calculated by maximizing a label assignment probability matrix (LAPM) with zero columns removed, which contains the information about the community structure at the group level. Each row in LAPM is a vector of label assignment probabilities (LAP) for a specific node. Finally, we model the FC of  brain networks. A Normal-NIG conjugate pair is used to model the connectivity $\boldsymbol{X}_{ij}=\lbrace x_{ij}^{1},x_{ij}^{2},\cdots,x_{ij}^{S}\rbrace$ from $S$ subjects for a specific pair of nodes. The mean and variance connectivity at the group level are estimated by drawing samples from the Normal (N) and Inverse Gamma (IG) posterior densities respectively.}
\label{Bayesian_Hierarchical}
\end{figure*}

We will first illustrate how to model the adjacency matrix of each subject with a latent block model and how to estimate individual-level community memberships using the MCMC sampling (Fig.\ref{Bayesian_Hierarchical}a). Then we will illustrate the details of using conjugate Bayesian pairs to model the estimated individual-level community memberships and individual adjacency matrices respectively. Finally, we perform parameter inference by drawing samples from posterior densities to depict the group representative network including the group-level community structure and group-level FC (Fig.\ref{Bayesian_Hierarchical}b).

\subsection{Individual-level modelling of community structure}
We first use a LBM \cite{Wyse2012,Bian2021} to model a weighted adjacency matrix $\boldsymbol{x}\in\mathbb{R}^{N\times N}$ measuring the FC of a specific brain network for each subject, where $N$ denotes the number of nodes which are assigned to $K$ communities. Suppose that $K$ is a random variable following a Poisson distribution $p(K)=\frac{\lambda^{K}}{K!}e^{-\lambda}$. Throughout this paper, we set the parameter $\lambda=1$ to simplify the distribution of $K$. We denote the community memberships as a latent label vector $\boldsymbol{z} = (z_{1}, \ldots,z_{i} ,\ldots, z_{N})$ where $z_{i}\in\lbrace 1,\cdots,K \rbrace$ is the label of node $i$. Each $z_{i}$ independently follows a categorical (one-trial multinomial) distribution: $z_{i}\sim \mbox{Categorical}(1; \boldsymbol{r}=(r_{1},\cdots,r_{k},\cdots,r_{K}))$, where $\boldsymbol{r}$ is a vector, the element $r_{k}$ is the probability of a node being assigned to community $k$, and $\sum_{k=1}^{K} r_{k}=1$. The categorical probability can be written using the indicator function $I_{k}(z_{i})$ as
\begin{equation}
p(z_{i}\vert\boldsymbol{r},K)=\prod_{k=1}^{K}r_{k}^{I_{k}(z_{i})}, \mbox{where\ }  I_{k}(z_{i})=
\begin{cases}
1, \ \mbox{if}\ z_{i}=k\\
0, \ \mbox{if}\ z_{i}\neq k\\
\end{cases}.
\end{equation}
The density of the $N$-dimensional latent label vector $\boldsymbol{z}$ is
\begin{equation}
p(\boldsymbol{z}\vert\boldsymbol{r},K)=\prod_{k=1}^{K}r_{k}^{m_k(\boldsymbol{z})},
\end{equation}
where $m_{k}(\boldsymbol{z})=\sum_{i=1}^{N}I_{k}(z_{i})$. The assignment probability vector $\boldsymbol{r}=(r_{1},\cdots,r_{k},\cdots,r_{K})$ follows a $K$-dimensional Dirichlet distribution 
$p(\boldsymbol{r}\vert K)=N(\boldsymbol{\alpha})\prod_{k=1}^{K}r_{k}^{\alpha_{k}-1}$,
where $N(\boldsymbol{\alpha})=\frac{\Gamma(\sum_{k=1}^{K}\alpha_{k})}{\prod_{k=1}^{K}\Gamma(\alpha_{k})}$ is the normalization factor. We assume that the community assignment is equally likely a priori before observing the data so that the hyperparameters are non-informative with $\alpha_k = 1$ for $k = 1, \ldots, K$ throughout the paper such that $\boldsymbol{r}$ follows the flat Dirichlet distribution with $p(\boldsymbol{r}\vert K)=(K-1)!$. 

We define the submatrix $\boldsymbol{x}_{kl}$ to be the weighted edges connecting the nodes in community $k$ to the nodes in community $l$, where $k,l\in \lbrace 1,\cdots,K\rbrace$. The likelihood of the LBM can be written as
\begin{equation}
p(\boldsymbol{x}\vert \boldsymbol{\pi},\boldsymbol{z},K)=\prod_{k.l}p(\boldsymbol{x}_{kl}\vert \pi_{kl},\boldsymbol{z},K),
\end{equation}
and the likelihood in specific blocks
\begin{equation}
p(\boldsymbol{x}_{kl}\vert \pi_{kl},\boldsymbol{z},K)=\prod_{\lbrace i\vert z_{i}=k \rbrace}\prod_{\lbrace j\vert z_{j}=l \rbrace} p(x_{ij}\vert \pi_{kl},\boldsymbol{z},K),
\end{equation}
where $\boldsymbol{\pi}=\lbrace \pi_{kl} \rbrace$ is a $K\times K$ model parameter matrix that is further characterized in the following section. 

\subsubsection{The latent block model with weighted edges}

For weighted edges, the block model parameter in block $kl$ consists of the block mean and variance $\pi_{kl}=(\mu_{kl},\sigma_{kl}^{2})$. Each $x_{ij}$ in the block $kl$ follows a Gaussian distribution conditional on the community number $K$ and the latent label vector $\boldsymbol{z}$, that is $x_{ij}\vert\pi_{kl}, \boldsymbol{z},K \sim \mathcal{N}(\mu_{kl},\sigma_{kl}^{2})$. The model parameter $\pi_{kl}=(\mu_{kl},\sigma_{kl}^{2})$ is assumed to independently follow the conjugate Normal-Inverse-Gamma (NIG) prior $\pi_{kl}\sim \mbox{NIG}(\xi,\kappa^2\sigma_{kl}^{2},\nu/2,\rho/2)$. That is $\mu_{kl}\sim\mathcal{N}(\xi,\kappa^2\sigma_{kl}^{2})$ and $\sigma_{kl}^{2}\sim \mbox{IG}(\nu/2,\rho/2)$. The density of the Inverse-Gamma distribution $\mbox{IG}(\alpha,\beta)$ has the general formula
$
p(x)=\frac{\beta^{\alpha}}{\Gamma(\alpha)}x^{-(\alpha+1)}e^{(\frac{-\beta}{x})}
$, where $\alpha$ and $\beta$ are hyperparameters.

We define $s_{kl}(\boldsymbol{x})$ to be the sum of the edge weights in the block $kl$: $s_{kl}(\boldsymbol{x})=\sum_{i:z_{i}=k}\sum_{j:z_{j}=l}x_{ij}$, and $q_{kl}(\boldsymbol{x})$ to be the sum of squares:
$q_{kl}(\boldsymbol{x})=\sum_{i:z_{i}=k}\sum_{j:z_{j}=l}x_{ij}^{2}$. We also define $w_{kl}(\boldsymbol{z})=m_{k}(\boldsymbol{z})m_{l}(\boldsymbol{z})$ to be the number of elements in the block, where $m_k$ and $m_l$ are the numbers of nodes in community $k$ and $l$ respectively. The prior and the likelihood constitute a NIG-Gaussian conjugate pair. With this conjugate pair, we can calculate the posterior distribution for each model block, which is also a Normal-Inverse-Gamma distribution 
$\mu_{kl}\sim\mathcal{N}(\xi_n,\kappa_n^2\sigma_{kl}^{2})$ and $\sigma_{kl}^{2}\sim \mbox{IG}(\nu_n/2,\rho_n/2)$, where $\nu_n=\nu+w_{kl}$, $\kappa_n^2=\frac{\kappa^2}{1+w_{kl}\kappa^2}$, $\xi_n=\frac{\xi+s_{kl}\kappa^2}{1+w_{kl}\kappa^2}$, and $\rho_n=\frac{\xi^{2}}{\kappa^{2}}+q_{kl}+\rho-\frac{(\xi+s_{kl}\kappa^{2})^2}{1/\kappa^{2}+w_{kl}}$. The detailed derivation can be found in Appendix A Supplementary materials (Section 2) of \cite{Bian2021}. The posterior density of the whole model is a product of such terms for all blocks, as follows.  
\begin{equation}
p(\boldsymbol{\pi}\vert\boldsymbol{x},\boldsymbol{z})=\prod_{k,l}p(\pi_{kl}\vert\boldsymbol{x}_{kl},\boldsymbol{z}).
\end{equation} 
Given a sampled $\boldsymbol{z}$ we can draw $\boldsymbol{\pi}$ from the above posterior directly.

\subsubsection{The collapsed posterior of the latent block model}
Our aim is to infer latent label vector $\boldsymbol{z}$ and the number of communities $K$ by drawing samples from the collapsed posterior $p(\boldsymbol{z}, K\vert \boldsymbol{x})$ \cite{ Wyse2012, MacDaid2012} which can be constructed by integrating out nuisance parameters. We start the derivation of the collapsing procedure with a joint density
\begin{equation}
p(\boldsymbol{x},\boldsymbol{\pi},\boldsymbol{z},\boldsymbol{r},K)=p(K)p(\boldsymbol{z},\boldsymbol{r}\vert K)p(\boldsymbol{x},\boldsymbol{\pi}\vert \boldsymbol{z}).
\end{equation}
The parameters $\boldsymbol{r}$ and $\boldsymbol{\pi}$ can be collapsed to obtain the marginal density $p(\boldsymbol{x},\boldsymbol{z},K)$ and the collapsed posterior $p(\boldsymbol{z}, K\vert \boldsymbol{x})$ is proportional to $p(\boldsymbol{x},\boldsymbol{z},K)$ as follows
\begin{align}
p(\boldsymbol{z},K, \boldsymbol{x})=p(K)\int p(\boldsymbol{z},\boldsymbol{r}\vert K)d\boldsymbol{r}  \prod_{k,l}\int p(\boldsymbol{x}_{kl},\pi_{kl}\vert \boldsymbol{z})d\pi_{kl}.
\end{align}
The first integral $p(\boldsymbol{z}\vert K)=\int p(\boldsymbol{z},\boldsymbol{r}\vert K)d\boldsymbol{r}$ is over the $K$-simplex $\{\boldsymbol{r}\! :\! \sum_{k=1}^{K} r_{k}=1\}$ and can be calculated as
\begin{align}
\int p(\boldsymbol{z},\boldsymbol{r}\vert K)d\boldsymbol{r}&=\frac{\Gamma(\sum_{k=1}^{K}\alpha_{k})}{\Gamma(\sum_{k=1}^{K}(\alpha_{k}+m_{k})}\prod_{k=1}^{K}\frac{\Gamma(\alpha_{k}+m_{k})}{\Gamma(\alpha_{k})},
\end{align}
while the second integral over $\pi_{kl}$ is
\begin{align}
\int p(\boldsymbol{x}_{kl},\pi_{kl}\vert \boldsymbol{z})d\pi_{kl}=\frac{\rho^{\nu/2}\Gamma\lbrace(w_{kl}+\nu)/2\rbrace}{\pi^{w_{kl}/2}\Gamma(\nu/2)(w_{kl}\kappa^2+1)^{1/2}}\nonumber
\\
\times(-\frac{\kappa^2(s_{kl}+\xi/\kappa^2)^2}{w_{kl}\kappa^2+1}+\frac{\xi^2}{\kappa^2}+q_{kl}+\rho)^{-(w_{kl}+\nu)/2}.
\end{align}
See the detailed derivation of the above two integrals in Appendix A Supplementary materials (Section 3 and 4) of \cite{Bian2021}.

\subsubsection{Estimation of community structure at the individual level}

Estimation of community structure at the individual level involves sampling a latent label vector $\boldsymbol{z}$ from the collapsed posterior distribution given the individual adjacency matrix $\boldsymbol{x}$ for each subject. There are several proposals for updating the latent label vector using the MCMC method, which constructs a chain of $\boldsymbol{z}$ samples converging to the collapsed posterior $p(\boldsymbol{z}\vert \boldsymbol{x},K)$. The strategies for updating $\boldsymbol{z}$ depend on whether $K$ is treated as a constant or a random variable. For invariant $K$, the number of communities is constant and possible updates include the Gibbs move and the M3 move. For variant $K$, the moves are the ejection and absorption moves, which can change the number of communities. In the ejection move, a community ejects another community,
and in the absorption move, a community absorbs another
community \cite{Nobile2007}. Combining absorption and ejection moves alone is difficult to converge during the run of Markov chain. Estimates of autocorrelation functions of absorption and ejection moves + Gibbs moves decay very slowly, while combining all these four moves decays faster. For individual-level inference, we integrate these four kinds of move into the Metropolis-Hastings algorithm \cite{Hastings1970}.

\subsubsection*{MCMC allocation sampler with invariant $K$}

We first elaborate on the update of the latent label vector $\boldsymbol{z}$ with proposal $p(\boldsymbol{z}\rightarrow\boldsymbol{z}^{\ast})$ \cite{Nobile2007} where $K$ is a fixed number. In the Metropolis-Hastings algorithm \cite{Hastings1970}, a candidate latent label vector $\boldsymbol{z}^{\ast}$ is accepted with probability $\min\lbrace 1,r \rbrace$, where
\begin{equation}
r=\frac{p(K,\boldsymbol{z}^{\ast},\boldsymbol{x})p(\boldsymbol{z}^{\ast}\rightarrow\boldsymbol{z})}{p(K,\boldsymbol{z},\boldsymbol{x})p(\boldsymbol{z}\rightarrow\boldsymbol{z}^{\ast})}.
\end{equation} 

\textbf{Gibbs move:} 
At each Gibbs move, one entry $z_{i}$ is randomly selected from $\boldsymbol{z}$ and updated by drawing from
\begin{equation}
p(z_{i}^{\ast}\vert \boldsymbol{z}_{-i},\boldsymbol{x},K)=\frac{1}{C} p(z_{1},\cdots,z_{i}^{\ast}=k,\cdots,z_{N}\vert\boldsymbol{x}),  
\end{equation}
where $k\in \lbrace 1,\cdots,K \rbrace$, and $\boldsymbol{z}_{-i}$ represents the elements in $\boldsymbol{z}$ apart from $z_{i}$. The normalization term can be written as
\begin{equation}
C=p(\boldsymbol{z}_{-i}\vert\boldsymbol{x},K)=\sum_{k=1}^{K}p(z_{1},\cdots,z_{i}^{\ast}=k,\cdots,z_{N}\vert\boldsymbol{x}).
\end{equation}
For the Metropolis-Hastings sampler with Gibbs move, the acceptance ratio is $r=1$. The computational complexity of a Gibbs move depends on the cost of calculating the probability of the reassignment of a specific entry. Each probability takes $O(K^2+N^2)$ time to calculate. There are $K$ possible reassignments so that each Gibbs move takes $O(K^3+KN^2)$ time.

\textbf{M3 move:} 
The M3 move can update multiple entries of the latent label vector $\boldsymbol{z}$. Two communities $k_{1}$ and $k_{2}$ are randomly selected in $\boldsymbol{z}$. We define a list $I=\lbrace i,z_{i}=k_{1} \mbox{\ or\ }z_{i}=k_{2}\rbrace$ with length $L$, and a vector of the labels apart from the list $\widetilde{\boldsymbol{z}}$. For the update, one element $z_{i}$ is randomly selected from the list and updated according to a reassignment probability $P_{k}^{i}$ as follows
\begin{eqnarray}
\frac{P_{k_{1}}^{i}}{P_{k_{2}}^{i}}&=&\frac{P_{k_{1}}^{i}}{1-P_{k_{1}}^{i}}\nonumber
\\
&=&\frac{p(z_{i}^{\ast}=k_{1},\widetilde{\boldsymbol{z}}\vert K)}{p(z_{i}^{\ast}=k_{2},\widetilde{\boldsymbol{z}}\vert K)}\frac{p(\widetilde{\boldsymbol{x}},\boldsymbol{x}^{\ast i}\vert z_{i}^{\ast}=k_{1},\widetilde{\boldsymbol{z}},K)}{p(\widetilde{\boldsymbol{x}},\boldsymbol{x}^{\ast i}\vert z_{i}^{\ast}=k_{2},\widetilde{\boldsymbol{z}},K)}.
\end{eqnarray}
Then, $z_{i}^{\ast}$ will be collected into $\widetilde{\boldsymbol{z}}$ at the next iteration.
The observation $\widetilde{\boldsymbol{x}}$ corresponds to $\widetilde{\boldsymbol{z}}$ and the observation $\boldsymbol{x}^{{\ast}i}$ corresponds to the updated $z_{i}^{\ast}$.
The ratio of the proposal can be written as
\begin{eqnarray}
\frac{p(\boldsymbol{z}^{\ast}\rightarrow\boldsymbol{z})}{p(\boldsymbol{z}\rightarrow\boldsymbol{z}^{\ast})}&=&\prod_{i\in I}\frac{P_{z_{i}}^{i}}{P_{z_{i}^{\ast}}^{i}}.
\end{eqnarray} For the detailed derivation of the M3 move, one can refer to Appendix A Supplementary materials (Section 5) in \cite{Bian2021}. The time complexity to calculate the ratio of posterior density is $O(K^2+N^2)$. The time complexity to calculate the ratio of proposals is $O(LN^2)$ depending on the length of the list, which changes dynamically during the update of the latent labels.

\subsubsection*{MCMC allocation sampler with variant $K$}

We can sample $K$ along with the latent label vector $\boldsymbol{z}$ from the collapsed posterior $p(\boldsymbol{z}, K \vert \boldsymbol{x})$ with proposal $p(\lbrace\boldsymbol{z}, K\rbrace\rightarrow\lbrace\boldsymbol{z}^{\ast}, K^{\ast}\rbrace)$ consisting of the absorption/ejection move \cite{Nobile2007} that changes $K$.
A candidate is accepted with probability $\min\lbrace 1,r \rbrace$, where
\begin{equation}
r=\frac{p(K^{\ast},\boldsymbol{z}^{\ast},\boldsymbol{x})p(\lbrace\boldsymbol{z}^{\ast}, K^{\ast}\rbrace\rightarrow\lbrace\boldsymbol{z}, K\rbrace)}{p(K,\boldsymbol{z},\boldsymbol{x})p(\lbrace\boldsymbol{z}, K\rbrace\rightarrow\lbrace\boldsymbol{z}^{\ast}, K^{\ast}\rbrace)}.
\end{equation} 
If $\lbrace\boldsymbol{z}, K\rbrace\rightarrow\lbrace\boldsymbol{z}^{\ast}, K^{\ast}\rbrace$ is the ejection move with acceptance probability $\min\lbrace 1,r \rbrace$, then the inverse move $\lbrace\boldsymbol{z}^{\ast}, K^{\ast}\rbrace\rightarrow\lbrace\boldsymbol{z}, K\rbrace$ is the absorption move with acceptance probability $\min\lbrace 1,\frac{1}{r}\rbrace$. Suppose that the maximum possible number of communities in the Markov chain is $K_{max}$ and we have $P_{K}^{E}$ chance to choose ejection move and $1-P_{K}^{E}$ chance to choose absorption move. If $K=K_{max}$ for the vector $\boldsymbol{z}$ at the current state, there must be an absorption move ($P_{K_{max}}^{E}=0$). For the current state of the latent label vector with $K=1$, there must be an ejection move ($P_{1}^{E}=1$). For $K=2,\cdots,K_{max}-1$, we set the probability of the ejection move as $P_{K}^{E}=0.5$ to have the same chance of choosing the ejection or the absorption move. In practical applications, users may adjust $K_{max}$ to regulate the size or number of communities, especially for networks that are expected to contain a larger number of communities.

\textbf{Ejection move:} We propose the ejection move $\lbrace\boldsymbol{z}, K\rbrace\rightarrow\lbrace\boldsymbol{z}^{\ast}, K^{\ast}\rbrace$, where $K^{\ast}=K+1$. For current state $\lbrace\boldsymbol{z}, K\rbrace$, the ejecting community $k_{1}$ is randomly selected from $K$ communities. The ejected community is labelled with $k_{2}=K+1$. The labels in the ejecting community are reassigned to $k_{1}$ with probability $p_{e}\sim \mbox{Beta}(u,u)$ or $k_{2}$ with probability $1-p_{e}$, and we set $u=1$ which makes it a noninformative uniform distribution. The proposal for the ejection move can be written as
\begin{align}
p(\lbrace\boldsymbol{z}, K\rbrace\rightarrow\lbrace\boldsymbol{z}^{\ast}, K^{\ast}\rbrace)=P_{K}^{E}\frac{1}{K}\frac{\Gamma(2u)}{\Gamma(u)\Gamma(u)}\nonumber
\\
\times\frac{\Gamma(u+\widetilde{m}_{k_{1}})\Gamma(u+\widetilde{m}_{k_{2}})}{\Gamma(2u+\widetilde{m}_{k_{1}}+\widetilde{m}_{k_{2}})},
\end{align}
where $\widetilde{m}_{k_{1}}$ and $\widetilde{m}_{k_{2}}$ are the numbers of elements reassigned to communities $k_{1}$ and $k_{2}$, respectively.

\textbf{Absorption move:} For the absorption move $\lbrace\boldsymbol{z}^{\ast}, K^{\ast}\rbrace\rightarrow\lbrace\boldsymbol{z}, K\rbrace$ where $K=K^{\ast}-1$, the absorbing community $k_{1}$ is randomly selected from the rest of $K$ communities and the absorbed community is $k_{2}=K^{\ast}$.
All the elements in $k_{2}$ are reassigned to $k_{1}$.
The proposal for the absorption move can be expressed as
\begin{align}
p(\lbrace\boldsymbol{z}^{\ast}, K^{\ast}\rbrace\rightarrow\lbrace\boldsymbol{z}, K\rbrace)=(1-P_{K}^{E})\frac{1}{K}.
\end{align}
The time complexity to calculate the ratio of posterior is $O(K^2+N^2)$, here $K$ is variant when updating the latent labels in the chain. The time complexity to calculate the ratio of proposals for the AE move is $O(N)$. 

In MCMC sampling for individual-level inference, we randomly select these four kinds of move with equal probability to update the latent label vector $\boldsymbol{z}$. The latent label vectors are sampled from the Markov chain after a predefined burn-in iteration to ensure the convergence of the chain, and with a specific autocorrelation time between the samples. In this paper, we set the burn-in iteration of 500, autocorrelation time of 3, and draw 400 samples of latent label vectors from the Markov chain. The sampled latent label vectors have the label switching problem, which is elaborated as follows.

\subsubsection*{Label switching}
Bayesian inference for community detection is subject to the label switching problem, a well-known issue in MCMC sampling for mixture models, where the labels in the latent assignment vector may interchange across iterations \cite{Stephens2000, Nobile2007, Wyse2012}. Additionally, the estimated latent labels may vary across individuals, leading to further inconsistencies as shown in Fig.\ref{Bayesian_Hierarchical}a. To address this issue, we employ a label switching method \cite{Stephens2000} that minimizes the distance between the latent label vectors, combined with a square assignment algorithm \cite{Carpaneto1980}, to relabel nodes and obtain a switched latent label vector.

In our latent block model, we set $\alpha_{k}=1$ with $\lbrace k=1,\cdots,K\rbrace$, and constant values of $\xi$, $\kappa^{2}$, $\nu$ and $\rho$ for the blocks $kl$, thus the prior remains invariant under permutations of the latent labels. Permutations of the latent labels do not change the likelihood, which means that the distributions with respect to blocks are not identifiable. Thus, the posterior distribution remains unchanged under permutations of latent labels.

We define a distance measure that quantifies the difference in coordinates between two latent label vectors $\boldsymbol{z}$ and $\boldsymbol{z}'$ as $D(\boldsymbol{z},\boldsymbol{z}')=\sum_{i=1}^{N}I(z_{i}\neq z'_{i})$, where $I$ is the indicator function.
We define
$\boldsymbol{\sigma}=\lbrace\sigma(1),\cdots,\sigma(k),\cdots,\sigma(K)\rbrace$
as a permutation of the labels $\lbrace 1,\cdots,k,\cdots, K\rbrace$.
We denote $\textbf{Q}=\lbrace\boldsymbol{z}^{j}(\boldsymbol{\sigma}^{j}), j=1,\cdots,J\rbrace$ as a collection of latent label vectors with respect to a sequence of permutations $\lbrace\boldsymbol{\sigma}^j, j=1,\cdots,J\rbrace$. We want to minimize the sum of all distances between the vectors $\sum_{j=1}^{J-1}\sum_{l=j+1}^{J} D(\boldsymbol{z}^{j}(\boldsymbol{\sigma}^{j}),\boldsymbol{z}^{l}(\boldsymbol{\sigma}^{l}))$. The solution to this minimization can be interpreted as a sequential optimization problem for square assignments. If the vectors that have already been relabelled up to $j-1$ are $\lbrace\boldsymbol{z}^{t},t=1,\cdots,j-1\rbrace$, for each vector $\boldsymbol{z}^{j}$ we define the element of a cost matrix $C(k_{1},k_{2})=\sum_{t=1}^{j-1}\sum_{i=1}^{N} D(z_{i}^{t}\neq k_{1},z_{i}^{j}= k_{2})$. We use the square assignment algorithm \cite{Carpaneto1980} that returns a permutation $\boldsymbol{\sigma}^{j}$, so that the total cost $\sum_{k=1}^{K}C(k,\sigma(k))$ for each $\boldsymbol{z}^{j}$ is minimized. Finally, the labels in the vector $\boldsymbol{z}^{j}$ are permuted according to $\boldsymbol{\sigma}^{j}$. The mode of the switched sampling $\boldsymbol{z}$ will be considered as the estimation of individual-level community memberships.

\subsection{Group-level modelling of the community structure and connectivity}
In this section, we model the community memberships which are estimated from the individual-level analysis. We consider the functional brain networks of all subjects and take individual community memberships (latent label vectors) as our observation for the group-level analysis (see in Fig.\ref{Bayesian_Hierarchical}b).

\subsubsection{Modelling community memberships}
For group-level analysis, we define a matrix 
\begin{equation}
\boldsymbol{Z} = 
\begin{pmatrix}
  z_{11} & \cdots & z_{1s} & \cdots & z_{1S}\\
  \vdots & \ddots & \vdots & \ddots & \vdots\\
  z_{i1} & \cdots & z_{is} & \cdots & z_{iS}\\
  \vdots & \ddots & \vdots & \ddots & \vdots\\
  z_{N1} & \cdots & z_{Ns} & \cdots & z_{NS}\\
\end{pmatrix}
\end{equation}
to represent the latent labels for all of the subjects estimated from the individual-level analysis, where $N$ is the number of nodes and $S$ is the number of subjects. The row vector $\boldsymbol{z}_{i}$ contains the labels of the group for a specific node $i$, and the column vector $\boldsymbol{z}_{s}$ represents the labels of a specific subject $s$ in the group. In group-level modelling, we model $\boldsymbol{z}_{i}=(z_{i1},\cdots,z_{is},\cdots,z_{iS})$ for $S$ subjects for node $i$ using a categorical-Dirichlet conjugate pair. Each label $z_{is}$ follows a categorical distribution $z_{is}\sim \mbox {Categorical}(1;\boldsymbol{r}_{i})$ where $\boldsymbol{r}_{i}=(r_{i1},\cdots,r_{ik},\cdots,r_{iK})$ is a vector of label assignment probabilities (LAP) such that $\sum_{k=1}^{K} r_{ik}=1$, and $K$ is the maximum element in $\boldsymbol{Z}$ in the group. We define a label assignment probability matrix (LAPM)
\begin{equation}
\boldsymbol{R}=\begin{pmatrix}
  r_{11} & \cdots & r_{1k} & \cdots & r_{1K}\\
  \vdots & \ddots & \vdots & \ddots & \vdots\\
  r_{i1} & \cdots & r_{ik} & \cdots & r_{iK}\\
  \vdots & \ddots & \vdots & \ddots & \vdots\\
  r_{N1} & \cdots & r_{Nk} & \cdots & r_{NK}\\
\end{pmatrix},
\end{equation}
and
$p(z_{is}\vert\boldsymbol{r}_{i},K)=\prod_{k=1}^{K}r_{k}^{I_{k}(z_{is})}$, where
$I_{k}(z_{is})=
\begin{cases}
1, \ \mbox{if}\ z_{is}=k\\
0, \ \mbox{if}\ z_{is}\neq k\\
\end{cases}$.
Consider a prior Dirichlet distribution $\boldsymbol{r}_{i}\sim \mbox{Dirichlet}(\boldsymbol{\alpha})$ with  $p(\boldsymbol{r}_{i}\vert K)=N(\boldsymbol{\alpha})\prod_{k=1}^{K}r_{k}^{\alpha_{k}-1}$, where the normalization factor $N(\boldsymbol{\alpha})=\frac{\Gamma(\sum_{k=1}^{K}\alpha_{k})}{\prod_{k=1}^{K}\Gamma(\alpha_{k})}$. The posterior can be formulated as
\begin{eqnarray}
p(\boldsymbol{r}_{i}\vert \boldsymbol{z}_{i},K)&\propto& \prod_{s=1}^{S} p({z}_{is}\vert \boldsymbol{r}_{i})p(\boldsymbol{r}_{i})\nonumber
\\
&=&\prod_{s=1}^{S}\prod_{k=1}^{K}r_{k}^{I_{k}(z_{is})}N(\boldsymbol{\alpha})\prod_{k=1}^{K}r_{k}^{\alpha_{k}-1}\nonumber
\\
&=& N(\boldsymbol{\alpha})\prod_{k=1}^{K}r_{k}^{\sum_{s=1}^{S}I_{k}(z_{is})+\alpha_{k}-1}\nonumber
\\
&=&\frac{N(\boldsymbol{\alpha})}{N(\boldsymbol{\alpha}')}N(\boldsymbol{\alpha}')\prod_{k=1}^{K}r_{k}^{\alpha_{k}'-1},
\end{eqnarray}
where $\alpha_{k}'=\sum_{s=1}^{S}I_{k}(z_{is})+\alpha_{k}$, and $N(\boldsymbol{\alpha}')=\frac{\Gamma(\sum_{k=1}^{K}(\alpha_{k}'))}{\prod_{k=1}^{K}\Gamma(\alpha_{k}')}$.
Therefore, we have the posterior $\boldsymbol{r}_{i}\vert \boldsymbol{z}_{i}\sim \mbox{Dirichlet}(\boldsymbol{\alpha}')$.
The posterior for the network can be expressed as $p(\boldsymbol{R}\vert \boldsymbol{Z},K)=\prod_{i=1}^{N}p(\boldsymbol{r}_{i}\vert \boldsymbol{z}_{i},K)$.

The nodes may have approximately similar probabilities to be allocated to two or multiple communities, in which case the communities are overlapped. It is a complicated question to evaluate the inter-individual variability about the community structure for overlapping communities. In this paper, we only explore within the scope of non-overlapping communities, which means that a node will only be allocated to a unique community. We use the maximum LAPM (MLAPM) $\boldsymbol{R}_{\mbox{max}}$, the maximum probability at each row of the matrix in $\boldsymbol{R}$, to provide the information about the community structure. Finally, all the zero columns are removed, resulting in a final version of MLAPM. The column index of the maximum probability in each row of MLAPM is the community label of the node.

\subsubsection{Modelling connectivity}
For modelling the connectivity of group representative network across subjects, We denote $\boldsymbol{X}_{ij}=(x_{ij}^{1},\cdots,x_{ij}^{s},\cdots,x_{ij}^{S})$ as a vector containing the connectivity between node $i$ and $j$ for $S$ subjects. We define a connectivity parameter for the element $ij$ as $\pi_{ij}=(\mu_{ij},\sigma_{ij}^{2})$ and a connectivity of each subject $x_{ij}^{s}$ follows a Gaussian distribution which is 
$x_{ij}^{s}\vert\pi_{ij} \sim \mathcal{N}(\mu_{ij},\sigma_{ij}^{2})$.

The connectivity parameter $\pi_{ij}=(\mu_{ij},\sigma_{ij}^{2})$ is assumed to independently follow the conjugate Normal-Inverse-Gamma (NIG) prior $\pi_{ij}\sim \mbox{NIG}(\xi,\kappa^2\sigma_{ij}^{2},\nu/2,\rho/2)$. That is $\mu_{ij}\sim\mathcal{N}(\xi,\kappa^2\sigma_{ij}^{2})$ and $\sigma_{ij}^{2}\sim \mbox{IG}(\nu/2,\rho/2)$. We define $w_{ij}(\boldsymbol{X}_{ij})$ to be the sum of the edge weights of connectivity $ij$ for the group and $q_{ij}(\boldsymbol{X}_{ij})$ to be the sum of squares as follows: $w_{ij}(\boldsymbol{X}_{ij})=\sum_{s=1}^{S}x_{ij}^{s}$, and $q_{ij}(\boldsymbol{X}_{ij})=\sum_{s=1}^{S}(x_{ij}^{s})^{2}$. The likelihood of the group connectivity model is
\begin{eqnarray}
p(\boldsymbol{X}_{ij}\vert \pi_{ij})
&=&\prod_{s=1}^{S}p(x_{ij}^{s}\vert \mu_{ij},\sigma_{ij}^{2}) \nonumber  \\
&=&(2\pi\sigma_{ij}^{2})^{-\frac{S}{2}}\mbox{exp}\lbrace -\frac{\sum_{s=l}^{S}(x_{ij}^{s}-\mu_{ij})^2}{2\sigma_{ij}^2}\rbrace \nonumber \\
&=&(2\pi\sigma_{ij}^{2})^{-\frac{S}{2}}\nonumber\\
&&\mbox{exp}\lbrace -\frac{q_{ij}-2\mu_{ij}w_{ij}+S\mu_{ij}^2}{2\sigma_{ij}^2}\rbrace.
\end{eqnarray}
With the Normal-NIG conjugate pair, we can calculate the posterior distribution for the group connectivity $\boldsymbol{X}_{ij}$, which is also a NIG distribution $\mbox{NIG}(\xi_s,\kappa_s^2\sigma_{ij}^{2},\nu_s/2,\rho_s/2)$. That is
$\mu_{ij}\sim\mathcal{N}(\xi_s,\kappa_s^2\sigma_{ij}^{2})$ and $\sigma_{ij}^{2}\sim \mbox{IG}(\nu_s/2,\rho_s/2)$. We derive the posterior of the model parameter $\pi_{ij}$ with prior $\mu_{ij}\sim\mathcal{N}(\xi,\kappa^2\sigma_{ij}^{2})$ and $\sigma_{ij}^{2}\sim \mbox{IG}(\nu/2,\rho/2)$ as follows.
\begin{eqnarray}
p(\pi_{ij}\vert\boldsymbol{X}_{ij})&\propto& p(\pi_{ij})p(\boldsymbol{X}_{ij}\vert\pi_{ij})\nonumber \\
&=& p(\mu_{ij})p(\sigma_{ij}^2)\prod_{s=1}^{S}p(x_{ij}^{s}\vert \mu_{ij},\sigma_{ij}^{2})\nonumber \\
&=&(2\pi\kappa^2\sigma_{ij}^2)^{-1/2}\nonumber \mbox{exp}\lbrace-\frac{(\mu_{ij}-\xi)^2}{2\kappa^2\sigma_{ij}^2} \rbrace\nonumber \frac{(\rho/2)^{\nu/2}}{\Gamma(\nu/2)}\\
&&\sigma_{ij}^{-2(\nu/2+1)}\mbox{exp}\lbrace-\rho/2\sigma_{ij}^2\rbrace\nonumber (2\pi\sigma_{ij}^{2})^{-S/2}\\
&&\mbox{exp}\lbrace -\frac{q_{ij}-2\mu_{ij}w_{ij}+S\mu_{ij}^2}{2\sigma_{ij}^2}\rbrace,
\end{eqnarray}
which can be rewritten as
\begin{eqnarray}
p(\pi_{ij}\vert\boldsymbol{X}_{ij})&\propto&\frac{(\rho/2)^{\nu/2}}{\Gamma(\nu/2)}(2\pi\kappa^2)^{-1/2}(2\pi)^{-S/2}\sigma_{ij}^{-1}\sigma_{ij}^{-\nu-2-S}\nonumber \\
&&\times\mbox{exp}\lbrace -\frac{1}{2\sigma_{ij}^2}[(\frac{1}{\kappa^2}+S)\mu_{ij}^2\nonumber \\
&&-2(\frac{1}{\kappa^2}\xi+w_{ij})\mu_{ij}+\frac{1}{\kappa^2}\xi^2+q_{ij}+\rho]\rbrace.
\end{eqnarray}
The posterior of the Gaussian model is also a Normal-Inverse-Gamma distribution which can be denoted as
$\mu_{ij}\sim\mathcal{N}(\xi_s,\kappa_s^2\sigma_{ij}^{2})$ and $\sigma_{ij}^{2}\sim \mbox{IG}(\nu_s/2,\rho_s/2)$. The posterior density can be expressed as
\begin{eqnarray}
p(\pi_{ij}\vert\boldsymbol{X}_{ij})&=&(2\pi\kappa_s^2\sigma_{ij}^2)^{-1/2}\mbox{exp}\lbrace-\frac{1}{2\kappa_s^2\sigma_{ij}^2}(\mu_{ij}-\xi_s)^2 \rbrace\nonumber \\
&&\times\frac{(\rho_s/2)^{\nu_s/2}}{\Gamma(\nu_s/2)}\sigma_{ij}^{-2(\nu_s/2+1)}\mbox{exp}\lbrace-\rho_s/2\sigma_{ij}^2\rbrace\nonumber \\
&=&\frac{(\rho_s/2)^{\nu_s/2}}{\Gamma(\nu_s/2)}(2\pi\kappa_s^2)^{-1/2}\sigma_{ij}^{-1}\sigma_{ij}^{-\nu_s-2}\nonumber \\
&&\times \mbox{exp}\lbrace -\frac{\frac{1}{\kappa_s^2}\mu_{ij}^2-\frac{2\xi_s}{\kappa_s^2}\mu_{ij}+\frac{\xi_s^2}{\kappa_s^2}+\rho_s}{2\sigma_{ij}^2}\rbrace.
\end{eqnarray}
Comparing the terms and coefficients with respect to $\mu_{ij}^2$, $\mu_{ij}$ and $\sigma_{ij}^2$, we have $-\nu_s-2=-\nu-2-S$, $\frac{1}{\kappa_s^2}=\frac{1}{\kappa^2}+S$, $\frac{2\xi_s}{\kappa_s^2}=2(\frac{1}{\kappa^2}\xi+w_{ij})$, and $\frac{\xi_s^2}{\kappa_s^2}+\rho_s=\frac{1}{\kappa^2}\xi^2+q_{ij}+\rho$. In summary, the parameters of the posterior density are given by $\nu_s=\nu+S$, $\kappa_s^2=\frac{\kappa^2}{1+S\kappa^2}$, $\xi_s=\frac{\xi+w_{ij}\kappa^2}{1+S\kappa^2}$, and $\rho_s=\frac{\xi^{2}}{\kappa^{2}}+q_{ij}+\rho-\frac{(\xi+w_{ij}\kappa^{2})^2}{1/\kappa^{2}+S}$. We can directly sample $\pi_{ij}$ from $\mbox{NIG}(\xi_s,\kappa_s^2\sigma_{ij}^{2},\nu_s/2,\rho_s/2)$.\\

\section{Experiments}
\subsection{Generative modelling and synthetic data experiments}
\subsubsection{Generative model}
To validate our hierarchical Bayesian modelling method, we propose a multivariate Gaussian generative model to simulate synthetic data of virtual subjects. Specifically, we generate $D$ segments of Gaussian time series. Within each segment, nodes are assigned to $K^{true}$ communities that differ in different segments. The true number of communities in the segments can be denoted as a vector $\boldsymbol{K}^{true}=\lbrace K_{1}^{true},\cdots,K_{d}^{true},\cdots,K_{D}^{true}\rbrace$.

We denote the true group-level label vectors in segments as $\lbrace \boldsymbol{z}_1,\cdots,\boldsymbol{z}_d,\cdots,\boldsymbol{z}_D \rbrace$. These are generated using the categorical-Dirichlet conjugate pair, i.e. the component weights $\lbrace \boldsymbol{r}_1, \cdots,\boldsymbol{r}_d,\cdots,\boldsymbol{r}_D \rbrace$ are first drawn from a uniform distribution on the $\boldsymbol{K}^{true}$ simplex and then nodes are assigned to the communities by drawing from the corresponding Categorical distributions. We also denote the true subject-specific label vectors $\lbrace \boldsymbol{z}_1^s, \cdots, \boldsymbol{z}_d^s,\cdots,\boldsymbol{z}_D^s \rbrace$ that determine the form of the covariance matrix in the generative model of the time series of a specific subject $s$. We introduce the degree of inter-individual variation (DIIV) which is an integer $n$ indicating the number of labels that are different between the baseline at the group level $\boldsymbol{z}_d$ and the subject-specific $\boldsymbol{z}_d^s$, so $n$ nodes are randomly chosen from a total of $N$ nodes, and each selected node is assigned a new label randomly drawn from the set $\lbrace 1,\cdots,K\rbrace$ for each subject. We set different levels of DIIV in simulation studies. The true subject-specific label vector $\boldsymbol{z}_d^s$ is used to generate synthetic data with the desired underlying community structure of each virtual subject. Specifically, time series data in $\mathbb{R}^{N}$ are simulated as
\begin{equation}
Y(t) =f(\boldsymbol{z}_d^s,a,b, t)+\boldsymbol{\epsilon}
\end{equation}
for $t = 1, \cdots, T$ by drawing $f(\boldsymbol{z}_d^s,a,b)\sim\mathcal{N}(\boldsymbol{0},\boldsymbol{\Sigma}_d^s(\boldsymbol{z}_d^s,a,b))$, with 
\begin{equation}
\Sigma_{ij}=\left\{
             \begin{array}{lr}
             1, \ \ \ \ \mbox{if}\ i\ = j\ \\
             a,\ \ \  \ \mbox{if}\ i\ \neq j\ \mbox{and}\ z_{i}=z_{j} \\    
             b,\ \ \  \ \mbox{if}\ i\ \neq j\ \mbox{and}\ z_{i}\neq z_{j}\\
             \end{array}
\right.
\end{equation}
where $a\sim U(\theta_{1},1)$ and $b\sim U(0,\theta_{2})$ are uniformly distributed and $\boldsymbol{\epsilon}\sim \mathcal{N}(\boldsymbol{0},\sigma^{2}\boldsymbol{I})$ is the additive Gaussian noise. Parameters $a$ and $b$ are proposed to control the correlation strength according to the community structure. Here, we set the uniform distribution parameters as $\theta_{1}=0.8$ and $\theta_{2}=0.2$, so that the nodes within the same community will have relatively stronger connectivity than those in different communities. The resulting covariance matrices for the $D$ segments are denoted as $\lbrace \boldsymbol{\Sigma}_{1}^s, \cdots,\boldsymbol{\Sigma}_{d}^s,\cdots,\boldsymbol{\Sigma}_{D}^s \rbrace$. For each virtual subject, the simulated Gaussian data $\boldsymbol{Y}^s\in \mathbb{R}^{N\times T}$ can be separated into $D$ segments which are denoted as $\lbrace \boldsymbol{Y}_1^s, \cdots,\boldsymbol{Y}_d^s,\cdots,\boldsymbol{Y}_D^s \rbrace$. 

For validation, we first generate 100 instances of synthetic data for a network with $N = 100$ nodes and $T=300$ time points. In the simulations, we generate $D=3$ data segments with a length of 100 frames for each segment corresponding to three experiments (Experiment 1, 2, and 3) and set the number of communities in the data segments to be $\boldsymbol{K}^{true}=\lbrace 8,9,10\rbrace$. The three segments of synthetic data under three experiments are simulated from the generative model with different levels of signal-to-noise ratio (SNR) from -20dB, -15dB, ... to 10dB for 100 instances. Here SNR $=10log_{10} (\Sigma_{ii}^{2}/\sigma^{2})$, and we set different values of $\sigma$ to control SNR. In addition, we set different levels of DIIV for $\lbrace \boldsymbol{z}_1^s, \cdots, \boldsymbol{z}_d^s,\cdots,\boldsymbol{z}_D^s \rbrace$ with DIIV=10, 20, and 30.

\subsubsection{Estimation evaluation}
Throughout this paper, we use NMI to evaluate the similarity between the switched estimated community memberships $\boldsymbol{z}$ and the true label vectors $\boldsymbol{z}_{true}$. We first define mutual information (MI) between switched estimation and ground truth as

\begin{equation}
I(\boldsymbol{z},\boldsymbol{z}_{true})=\sum_{k\in\boldsymbol{z}}\sum_{l\in\boldsymbol{z}_{true}}P(k,l)\mbox{log}(\frac{P(k,l)}{P(k)P(l)}),   
\end{equation}
where $P(k,l)=\frac{n_{kl}}{N}$ is the joint probability of a node in community $k$ in vector $\boldsymbol{z}$ and community $l$ in vector $\boldsymbol{z}_{true}$, $n_{kl}$ is the number of nodes shared between community $k$ in $\boldsymbol{z}$ and community $l$ in $\boldsymbol{z}_{true}$, $P(k)=\frac{n_{k}}{N}$ and $P(l)=\frac{n_{l}}{N}$ are the marginal probabilities with $n_{k}$ and $n_{l}$ as the size of communities in $\boldsymbol{z}$ and $\boldsymbol{z}_{true}$ respectively, and $N$ is the total number of nodes. We also denote the entropy of the partitions of $\boldsymbol{z}$ and $\boldsymbol{z}_{true}$ as 
$H(\boldsymbol{z})=-\sum_{k\in\boldsymbol{z}}P(k)\mbox{log}P(k)$, and $H(\boldsymbol{z}_{true})=-\sum_{l\in\boldsymbol{z}_{true}}P(l)\mbox{log}P(l)$ respectively. Then, NMI is calculated as
\begin{equation}
    \mbox{NMI}(\boldsymbol{z},\boldsymbol{z}_{true})=\frac{I(\boldsymbol{z},\boldsymbol{z}_{true})}{\sqrt{H(\boldsymbol{z})H(\boldsymbol{z}_{true})}}.
\end{equation}


\subsection{Task fMRI data experiments}
\subsubsection{Task paradigm}

In N-back working memory tasks, pictures of faces, places, tools, and body parts were shown in front of participants in each block. In the 2-back condition, the participants judged whether the current stimulus was the same as the one presented two steps earlier. In 0-back blocks, the participants judged whether any stimulus was the same as the target cue at the beginning of the block. There were 405 frames (TR of 0.72 s) with 4 blocks of 2-back, 4 blocks of 0-back conditions (each lasting 25 s) and 4 blocks for fixation (each lasting 15 s) in the paradigm of working memory task fMRI, including sequential blocks of (1) 2-back Tool, (2) 0-back Body, (3) Fixation, (4) 2-back Face, (5) 0-back Tool, (6) Fixation, (7) 2-back Body, (8) 2-back Place, (9) Fixation, (10) 0-back Face, (11) 0-back Place, and (12) Fixation. 

\subsubsection{Task fMRI data acquisition and preprocessing}

The working memory task fMRI data from 100 unrelated healthy subjects were collected and released by the Human Connectome Project (HCP) \cite{Barch2013}. No additional institutional review board (IRB) approval was required and informed consent was provided by all participants. The whole-brain echo-planar imaging (EPI) was acquired with a 32-channel head coil on a modified 3T Siemens Skyra (TR = 0.72 s, TE = 33.1 ms, flip angle = 52 degrees, BW = 2290 Hz/Px, in-plane FOV = $208\times 180$ mm, 72 slices with isotropic voxels of 2 mm with a multi-band acceleration factor of 8). The task fMRI includes two runs (left to right (LR) and right to left (RL)). The task fMRI data in the HCP dataset were minimally preprocessed with a pipeline based on FSL (FMRIB's Software Library) \cite{smith2004}.

\subsubsection{Time series extraction and brain network construction}
We extracted the BOLD signals of 100 brain regions of interest (ROIs) using Schaefer's 7-network atlas \cite{Schaefer2018} and those of 200 ROIs using Kong's 17-network atlas \cite{Kong2019}. The functional brain networks were constructed by calculating the Pearson's correlation coefficient of the extracted time series.

\subsection{Hyperparameter optimization}

For comparing the performance of our proposed latent block model with the commonly used (multilayer) modularity models, the hyperparameters of each model were optimized independently using Bayesian optimization.

For latent block model, the hyperparameters $\nu$, $\rho$, $\kappa^{2}$, and $K_{max}$ were optimized. The hyperparameter $\xi$ was set to 0 during optimization, thereby ensuring symmetry of $\mu_{kl}$ about 0 while maintaining weakly informative priors without increasing model complexity. The search space of $\nu$ was set as $\nu\in [2.1, 15]$. Setting the lower bound slightly above 2 ensures a well-defined variance while allowing heavy-tailed behavior. Larger values of $\nu$ increasingly concentrate the variance prior and approach Gaussian-like behavior. Empirically, values beyond approximately 15 produce limited additional regularization while reducing flexibility. Therefore, 15 serves as a practical upper limit that maintains model adaptivity. We set $\rho\in[0.001,1]$, where the lower bound 0.001 supports large variance values when supported by data and the upper bound 1 prevents excessive shrinkage of variance parameters. The hyperparameter $\kappa^{2}$ controls the prior variance of block means, and we set it as $\kappa^{2}\in [0.1,20]$. The lower bound 0.1 prevents over concentration around the prior mean with $\xi$=0, which would excessively constrain block means. Upper bound of 20 allows substantial dispersion in block means without introducing instability in posterior updates. We set $K_{max}\in [6,25]$, fewer than six communities is unlikely to capture known large-scale brain network organization. The upper bound 25 prevents over fragmentation into excessively small or noise driven communities. 

In modularity model, there is a resolution parameter $\gamma$ that controls the number and size of communities. After manual calibration, we set $\gamma \in [1,2.5]$ for synthetic data and $\gamma \in [1.5,3.5]$ for empirical data to ensure that the local minima identified by Bayesian optimization were contained within the predefined search intervals. In multilayer modularity model, apart from the modularity resolution $\gamma$, there is another parameter $\omega$ that characterizes the coupling of the node in one network to itself in another network.
For high values of $\omega$, the multilayer modularity approach may fail to capture inter-individual variability in subject-specific community assignments. We set $\gamma \in [1,2.5]$ and $\omega \in [0,0.1]$ for synthetic data and $\gamma \in [2, 4]$ and $\omega \in [0,0.1]$ for real data to ensure that the local minima of the multilayer modularity objective were contained within the specified intervals.

To formalize the objective function, let $\boldsymbol\theta$ denote the hyperparameter vector. For a given $\boldsymbol\theta$, community memberships were estimated for a group of subjects based on latent block model ($\boldsymbol\theta$=[$\nu$, $\rho$, $\kappa^{2}$, $K_{max}$]), modularity ($\boldsymbol\theta$=$\gamma$), or multilayer modularity ($\boldsymbol\theta$=[$\gamma$, $\omega$]). For Bayesian optimization using synthetic data, the objective function was defined as $f(\boldsymbol\theta)=-\frac{1}{S}\sum_{s=1}^{S}\mbox{NMI}(\boldsymbol{z}^{s}(\boldsymbol\theta),\boldsymbol{z}_{true}^{s})$ for $S$ subjects, where $\boldsymbol{z}^{s}(\boldsymbol\theta)$ is the estimated community memberships and $\boldsymbol{z}_{true}^{s}$ is the true subject-specific community memberships for subject $s$. The optimization problem is $\boldsymbol\theta^{\ast}=\mbox{arg} \ \mbox{min}_{\boldsymbol\theta}(f(\boldsymbol\theta))$. For real working memory task fMRI data, four FC matrices corresponding to four task conditions with the same working memory load (e.g., four 2-back conditions) were extracted and jointly evaluated for each subject. For subject $s$, this yielded four latent label vectors $\boldsymbol{z}_{1}^s$, $\boldsymbol{z}_{2}^s$,  $\boldsymbol{z}_{3}^s$ and  $\boldsymbol{z}_{4}^s$. For LBM and modularity, community detection was performed separately on the FC matrix of each subject. In contrast, multilayer modularity was applied to the FC matrices of all subjects jointly. To quantify within-subject consistency across task conditions, we split the four task conditions into two halves and computed normalized mutual information (NMI) between paired latent label vectors (split-half reproducibility). The objective function is $f(\boldsymbol{\theta})=- \mbox{corr} (\lbrace \mbox{NMI}(\boldsymbol{z}_{1}^s,\boldsymbol{z}_{2}^s)\rbrace_{s=1}^{S},\lbrace\mbox{NMI}(\boldsymbol{z}_{3}^s,\boldsymbol{z}_{4}^s) \rbrace_{s=1}^{S})$. Optimization was performed for 100 evaluations, and the parameter set minimizing the objective function was selected for subsequent analyses.

\section{Results}

In this section, we validate our method by demonstrating the results of analyzing both synthetic and real data, including individual-level and group-level validation. Our proposed individual-level community detection method with latent block model is compared with the commonly used modularity \cite{Newman2004, Newman2006, Bian2025} and multilayer modularity \cite{Mucha2010, Bassett2011,Bassett2013}. Our group-level community detection method with the Categorical-Dirichlet (CD) conjugate pair model is compared with the commonly used majority-voting (MV) strategy. We also validate the performance of group-level functional connectivity estimation by evaluating the similarity between posterior mean and real mean, and between posterior variance and real variance, respectively.

\subsection{Synthetic data analysis}

\subsubsection{Individual-level validation}

\begin{figure}[!ht]
\centering
\includegraphics[width=1\linewidth]{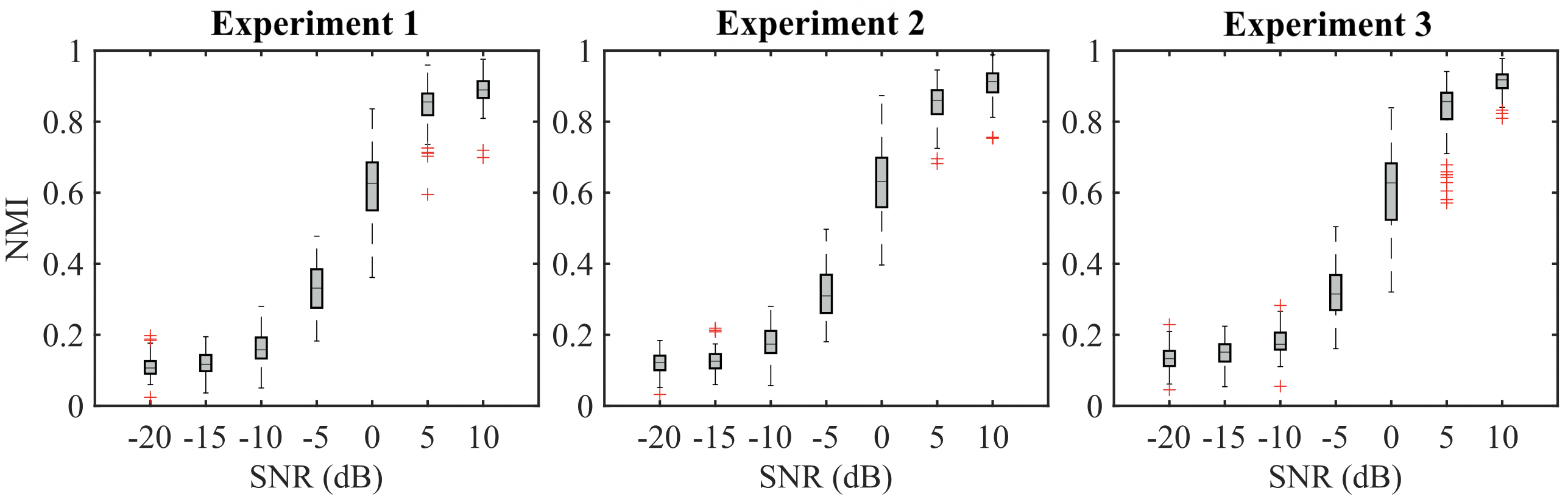}
\caption{\footnotesize The NMI between estimated individual-level community memberships and subject-specific true community memberships against different levels of SNR based on synthetic data with DIIV=10 using LBM.}
\label{Stat_analy_LBM_vs_SNR}
\end{figure}

We first evaluate the individual-level community detection performance of LBM using synthetic data with different levels of SNR from SNR = -20dB, -15dB to 10dB. Here we fix the DIIV to be 10. We set the hyperparameters as $[\nu$, $\rho$, $\kappa^{2}$, $K_{max}]=[3, 0.02, 1, 20]$ in this experiment, so that the prior Normal and IG distributions are neither too sharp nor too spread. For each virtual subject, we calculate the NMI between the estimated vector of community memberships and the true subject-specific community memberships, the results of which are shown in Fig.\ref{Stat_analy_LBM_vs_SNR}.

\begin{figure}[!ht]
\centering
\includegraphics[width=1\linewidth]{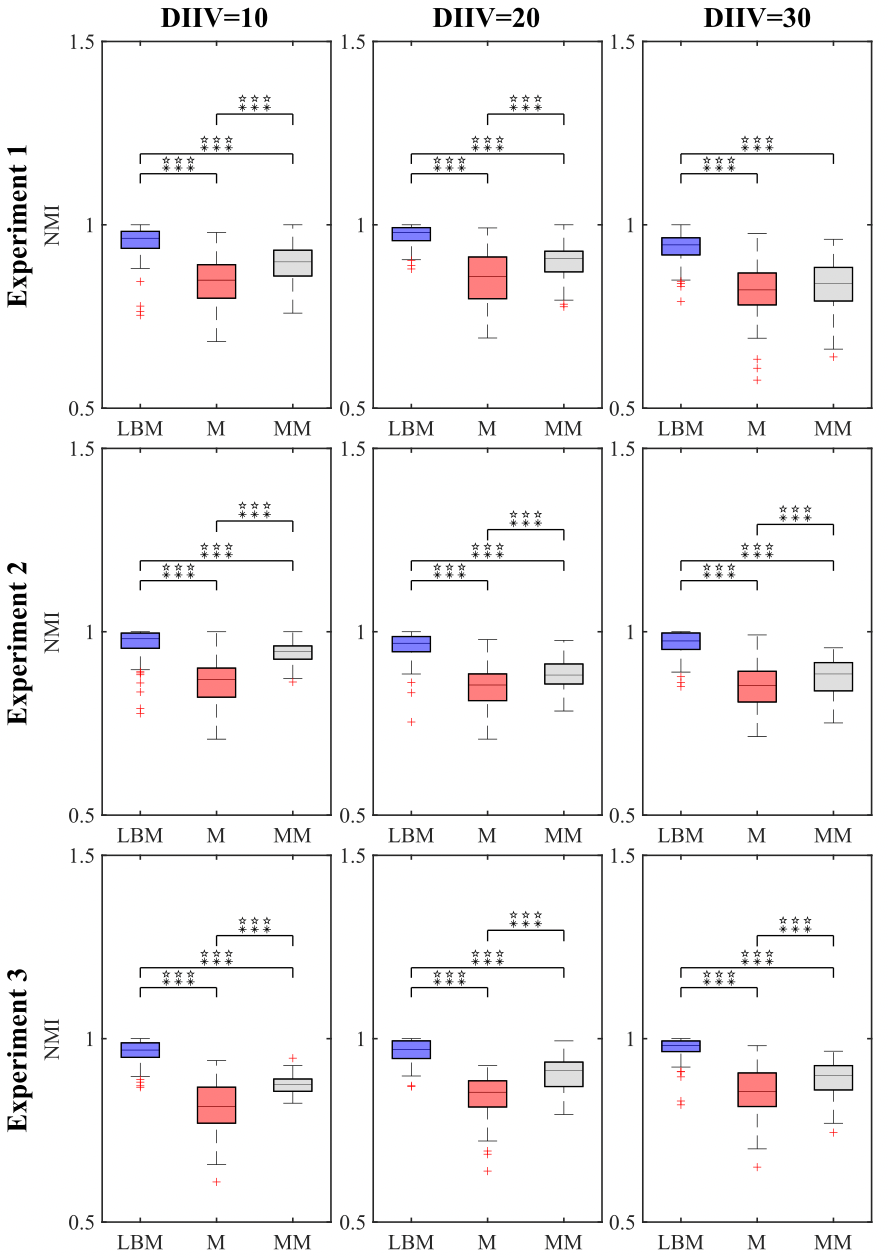}
\caption{\footnotesize Statistical comparisons between the performance of latent block model (LBM), modularity (M), and multilayer modularity (MM) for individual-level community detection using synthetic data of 10dB. The hyperparameters of each model were optimized using Bayesian optimization. Normalized mutual information (NMI) is evaluated between the estimated individual-level latent labels and the subject-specific true labels with different values of degree of inter-individual variation (DIIV). Experiment 1, 2, and 3 correspond to synthetic data segments 1, 2, and 3, respectively. The star denotes the significance level of the F-test for variance homogeneity (\ding{73}: $0.01\leq p<0.05$, \ding{73}\ding{73}: $0.001\leq p<0.01$, and \ding{73}\ding{73}\ding{73}: $p<0.001$). If the p-value of F-test is smaller than 0.05, Welch's t-test is used, otherwise, conventional two sample t-test is used. The asterisk indicates the statistical significance of t-test ($\ast$: $0.01\leq p<0.05$, $\ast$ $\ast$: $0.001\leq p<0.01$, and $\ast$ $\ast$ $\ast$: $p<0.001$). The p-values were corrected for multiple comparisons using the Benjamini-Hochberg false discovery rate (FDR) procedure.}
\label{Stat_analy_LBM}
\end{figure}

Next, we compare the individual-level community detection performance of LBM with the commonly used modularity and multilayer modularity models. For fair comparision, the hyperparameters of each model were optimised based on Bayesian optimization. According to the results in Fig.\ref{Stat_analy_LBM}, the NMI between the estimated individual-level community memberships and the ground truth of using LBM is statistically larger than that of using (multilayer) modularity with different levels of DIIV and experiments. These results indicate that LBM can detect community structure more accurately than (multilayer) modularity models.

\subsubsection{group-level validation}

\begin{figure}[!ht]
\centering
\includegraphics[width=1\linewidth]{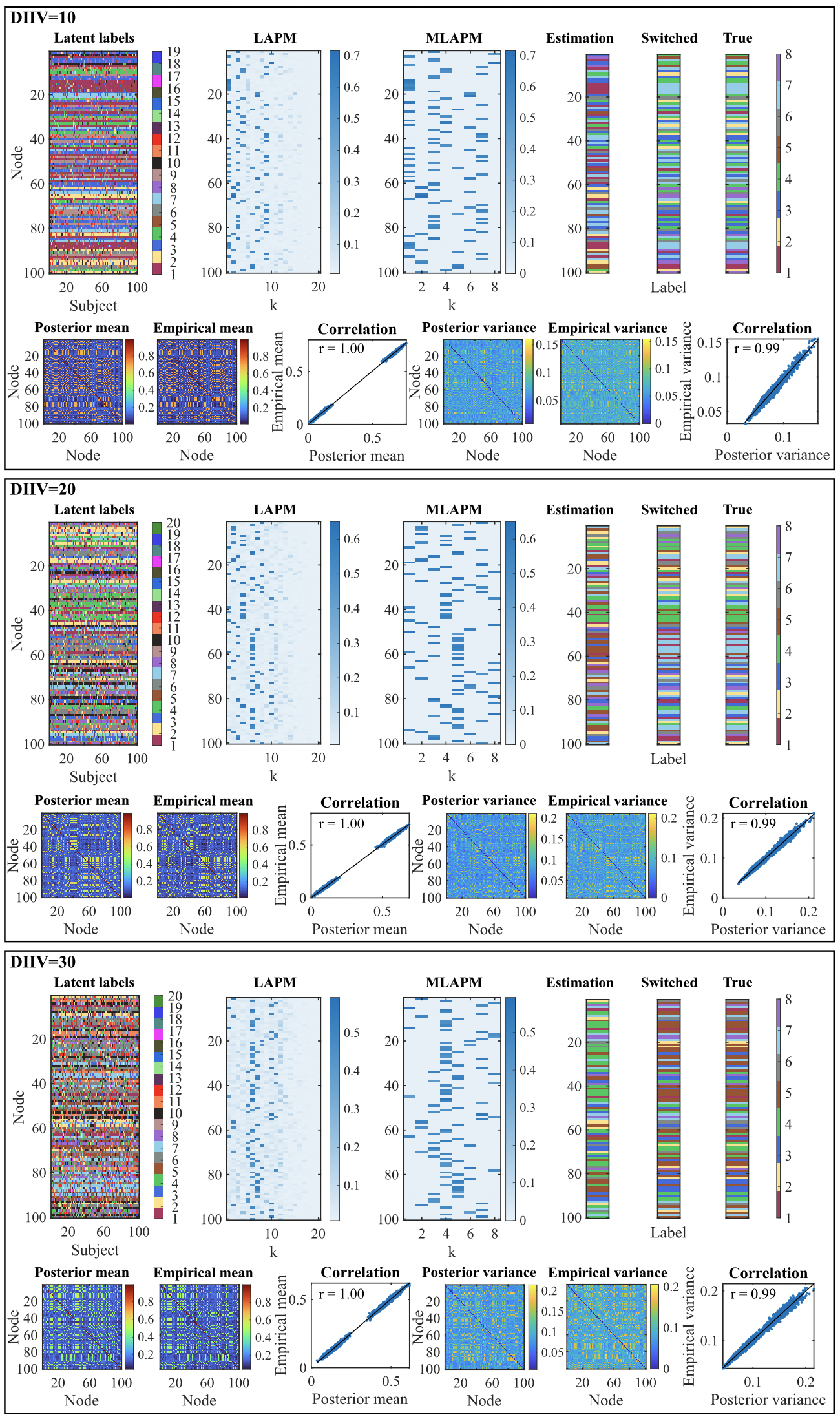}
\caption{\footnotesize Validation of the estimates of group-level community memberships and connectivity using different degree of inter-individual variation (DIIV) with SNR of 10dB for Experiment 1.}
\label{figure_synthetic}
\end{figure}

Individual-level latent labels estimated using LBM with an SNR of 10dB are shown in different colors in the leftmost panels of Fig.\ref{figure_synthetic} ($[\nu$, $\rho$, $\kappa^{2}$, $K_{max}]=[3, 0.02, 1, 20]$). A row ($\boldsymbol{z}_{i}$) represents the labels of 100 subjects for a specific node, and a column ($\boldsymbol{z}_{s}$) indicates the network labels of a subject. 

At the group level, categorical-Dirichlet conjugate pairs are used to model the latent labels in $\boldsymbol{Z}$ estimated at the individual level. An LAPM is estimated by drawing samples from a Dirichlet posterior density in the group-level analysis as shown in the second columns in the panels. Only the largest probability in each row of the LAPM is retained, and all zero columns are removed, resulting in an MLAPM. The column index of the MLAPM corresponds to the community memberships. The color bars indicate the community assignment probability in the LAPM and the MLAPM. The column indices in MLAPM are then transformed into a vector of labels and a label-switching algorithm \cite{Stephens2000} is applied. The number of columns that contain the maximum assignment probabilities indicates the number of communities. These estimates closely align with the true label vector used in the generative model, which is shown on the right-most side of each panel. We find that the performance of the community detection at the group level is not affected by DIIV (also see TABLE \ref{Table_DIIV}).

\begin{table}[h]
\centering
\renewcommand{\arraystretch}{1.25}
\caption{NMI between group-level estimated latent labels and group-level true labels against different levels of DIIV using LBM based on synthetic data with SNR=10dB}
\normalsize
{\tiny
\resizebox{\linewidth}{!}{
\begin{tabular}{l|ccc}
    \Xhline{1pt}
    \multirow{2}{*}{\textbf{DIIV}} & \multicolumn{3}{c}{\textbf{NMI}}  \\ \cline{2-4}
     & Experiment 1 & Experiment 2 & Experiment 3\\
    \hline
    10 & 1.0000 & 1.0000 & 1.0000
    \\
    20 & 0.9866 & 0.9915 & 0.9632
    \\
    30 & 1.0000 & 0.9677 & 0.8834
    \\
    \Xhline{1pt}
\end{tabular}
}
}
\label{Table_DIIV}
\end{table}

\begin{figure}[!ht]
\centering
\includegraphics[width=1\linewidth]{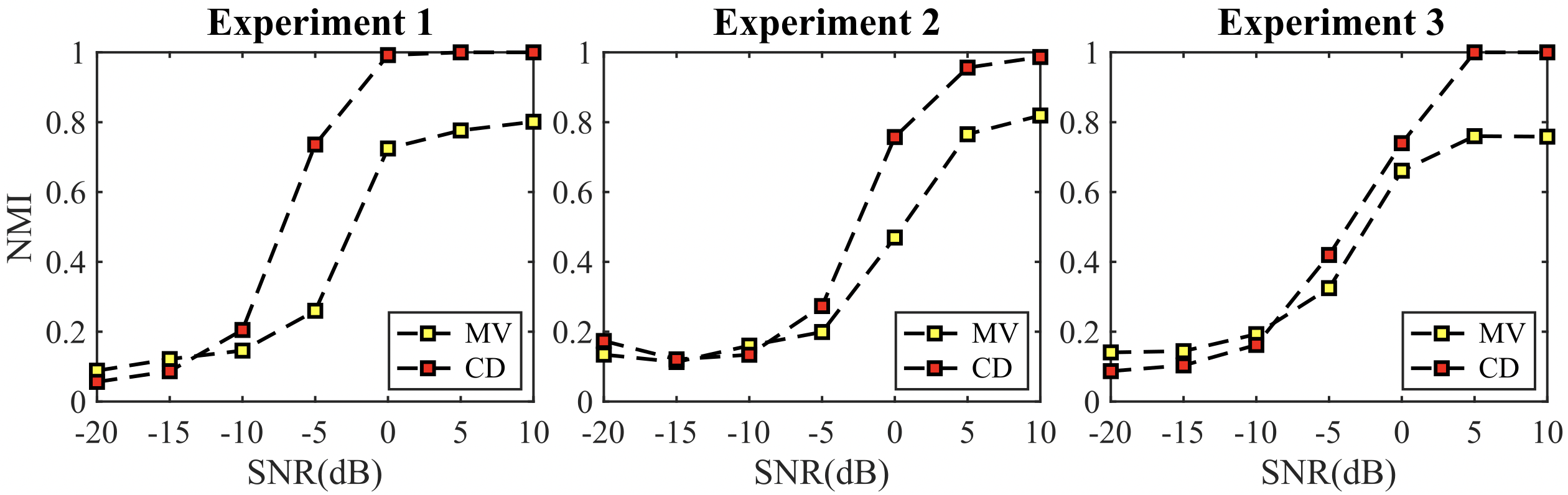}
\caption{\footnotesize Group-level MNI against SNR with DIIV=10 (MV: majority vote and CD: categorical-Dirichlet conjugate pair).}
\label{Group_LBM_vs_SNR}
\end{figure}

We also evaluate the performance of group-level community detection using a categorical-dirichlet conjugate pair (CD) with different levels of SNR and compare it with conventional majority voting (MV) strategy commonly used in machine learning, as shown in Fig.\ref{Group_LBM_vs_SNR}. Our CD method outperforms the MV for signals with SNR greater than  -5dB.

Next, we show the results of modelling the individual-level adjacency matrices using the Normal-Normal-Inverse-Gamma conjugate pair (Normal-NIG) with the parameter setting $[\nu$, $\rho$, $\kappa^{2}$, $K_{max}]=[3, 0.02, 1, 20]$. We compare the posterior sample with the real mean and variance of connectivity between a specific pair of nodes $i$ and $j$ using 100 virtual subjects to validate the precision of the proposed generative model and the Bayesian inference algorithm. We found that the posterior samples of mean and variance were highly correlated with the empirical mean and variance, and the Pearson's correlation coefficient values are shown in Fig.\ref{figure_synthetic}. The dots in the plot are divided into two groups for posterior samples of the mean. This is because the elements $a$ of the covariance matrix in the generative model indicating intra-community couplings are sampled from $a\sim U(0.8,1)$ and the elements $b$ indicating inter-community couplings are sampled from $b\sim U(0,0.2)$. The elements $a$ and $b$ determine the density of blocks in the adjacency matrix that shows a prominent discrepancy between intra-community connectivity and inter-community connectivity. Higher DIIV values result in more spread mean samples and larger values of variance samples.

\subsection{Working memory task fMRI data analysis}

\begin{figure*}[!ht]
\centering
\includegraphics[width=1\linewidth]{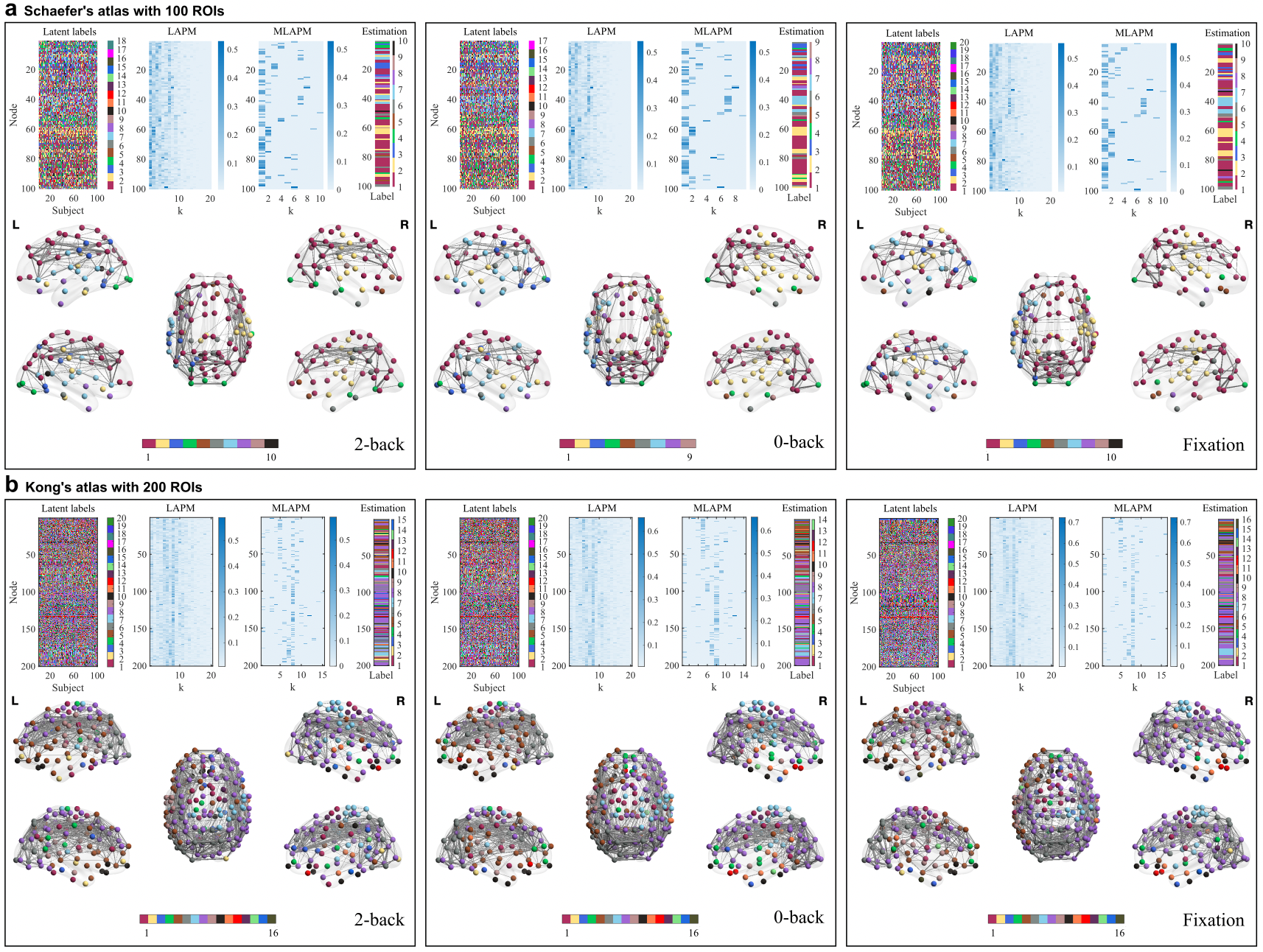}
\caption{\footnotesize The results of community detection and connectivity using working memory task fMRI data. \textbf{a} Schaefer's atlas with 100 ROIs. \textbf{b} Kong's atlas with 200 ROIs. We show the results of 2-back, 0-back, and fixation. The matrix at the upper left in each panel is the estimation of community memberships at individual level. Different colors represent community memberships of functional brain networks for 100 unrelated healthy subjects. The rows correspond to the nodes and the columns correspond to the subjects. The calculations of LAPM and MLAPM are the same to those in Fig. \ref{figure_synthetic}  The different colors of the labels represent community memberships which are the column indices in the MLAPM for each node. We visualize the brain networks using the BrainNet Viewer with a sparsity level of 5\%. }
\label{figure_real_Schaefer_Kong}
\end{figure*}

\begin{table}[h]
\centering
\renewcommand{\arraystretch}{1.25}
\caption{The split-half reproducibility of latent block model (LBM), modularity (M), and multilayer modularity (MM) using Schaefer's atlas and Kong's atlas in working memory task fMRI, evaluated by the correlation coefficient $r$ of the values of NMI of the two splits.}
\resizebox{\linewidth}{!}
{\begin{tabular}{cc|ccc|ccc}
    \Xhline{1pt}
       \textbf{1st half}& \textbf{2nd half}& \multicolumn{3}{c}{\textbf{Schaefer's atlas}}  \vline & \multicolumn{3}{c}{\textbf{Kong's atlas} }
     \\
    \Xhline{1pt}
    \multicolumn{2}{c}{\textbf{2-back}} \vline &
    \textbf{LBM} & \textbf{M} & \textbf{MM} & \textbf{LBM} & \textbf{M} & \textbf{MM}  
     \\
    \hline
    T $\&$ F & B $\&$ P & 0.6792 & 0.3881 & 0.4455 & 0.7497 & 0.4319 & 0.4485 
    \\
    T $\&$ B & F $\&$ P & 0.6794 & 0.3863 & 0.4588 & 0.7354 & 0.3902 & 0.3794 
    \\
    T $\&$ P & F $\&$ B & 0.7161 & 0.4515 & 0.5261 & 0.6942 & 0.4848 & 0.5403 
    \\
    \hline
    \multicolumn{2}{c}{Mean} \vline& 0.6916 & 0.4086 & 0.4768 & 0.7264 & 0.4356 & 0.4561
    \\
    \Xhline{1pt}
    \multicolumn{2}{c}{\textbf{0-back}} \vline&
    \textbf{LBM} & \textbf{M} & \textbf{MM} & \textbf{LBM} & \textbf{M} & \textbf{MM} 
     \\
    \hline
    T $\&$ F & B $\&$ P & 0.5685 & 0.3738 & 0.5399 & 0.6478 & 0.3350 & 0.4763 
    \\
    T $\&$ B & F $\&$ P & 0.5693 & 0.4559 & 0.4245 & 0.6242 & 0.5051 & 0.4150  
    \\
    T $\&$ P & F $\&$ B & 0.6230 & 0.3745 & 0.3748 & 0.6071 & 0.4282 & 0.3589  
    \\
    \hline
    \multicolumn{2}{c}{Mean} \vline & 0.5869 & 0.4014 & 0.4464 & 0.6264 & 0.4228 & 0.4167
    \\
    \Xhline{1pt}
\end{tabular}}
\label{Table_split_half_reproducibility}
\end{table}

\subsubsection{Individual-level validation}
To validate the performance of LBM, modularity and multilayer modularity using working memory task fMRI data, we use split-half reproducibility. As before, the hyperparameters of each model were optimized using Bayesian optimization. For reproducibility analysis, the four types of 2-back or 0-back conditions were divided into two halves. Three combinations of splits were used: (i) Tool $\&$ Face (T $\&$ F) in the first half and Body $\&$ Place (B $\&$ P) in the second half, (ii) Tool $\&$ Body (T $\&$ B) in the first half and Face $\&$ Place (F $\&$ P) in the second half, and (iii) Tool $\&$ Place (T $\&$ P) in the first half and Face $\&$ Body (F $\&$ B) in the second half. We first compute the NMI between conditions within the first and second halves, respectively, and then calculate the Pearson's correlation coefficient between the resulting NMI scores. Higher Pearson correlation values indicate greater reproducibility and reliability. The split-half reproducibility results for LBM, modularity, and multilayer modularity are presented in TABLE \ref{Table_split_half_reproducibility}. Overall, LBM demonstrates higher reproducibility and reliability compared to both modularity and multilayer modularity.

\subsubsection{Group-level validation}

For working memory task fMRI, we show the results of analyzing the data from LR phase encoding using LBM ($[\nu$, $\rho$, $\kappa^{2}$, $K_{max}]=[3, 0.02, 1, 20]$). The locations of the brain network architectures under the task of working memory are illustrated in previous work \cite{Bian2021}. Typical three brain network architectures corresponding to the first three experimental conditions of the paradigm 2-back, 0-back, and fixation are shown in Fig.\ref{figure_real_Schaefer_Kong}. Estimated labels (i.e., community memberships of nodes) of individual-level analysis for 100 unrelated healthy subjects are shown to the top left in each panel of Fig.\ref{figure_real_Schaefer_Kong} based on Schaefer's atlas with 100 ROIs and Kong's atlas with 200 ROIs. Different colors represent the labels of the community memberships. These labels are the estimation at the individual level and are considered as the observation at the group-level analysis. The group-level community structure is estimated by calculating the MLAPM matrix, where the number of columns represents the estimated number of communities $K$ at the group level. The row index indicates the node number and the column index indicates the community memberships at the group level. The bar shows the value of the maximum assignment probability of the labels.  The community memberships of brain networks under different experimental conditions are inconsistent with each other due to the label-switching phenomenon. Here, we used the relabelling algorithm \cite{Bian2021} to reassign the labels across different conditions. Brain network connectivity is visualized using BrainNet Viewer \cite{Xia2013}. The group-level weighted edges are at a sparsity level of 5\%, which is the mean connectivity estimated by drawing samples from the posterior density $\mu_{ij}\sim\mathcal{N}(\xi_s,\kappa_s^2\sigma_{ij}^{2})$. 

According to the results of Schaefer's atlas with 100 ROIs, we observe that nodes within the frontoparietal cortex are more cohesively organized into a single module (red module k=1) with stronger intra-module connectivity during the 2-back condition compared to the 0-back and fixation conditions, suggesting greater engagement of the frontoparietal network during higher working memory load. In contrast, the 0-back condition shows stronger integration within a sensory-motor and attention-related module (yellow module k=2). Additionally, the default mode network is more suppressed during the 2-back condition, as indicated by a smaller and less distinct default mode network (blue module k=7) compared to that observed in the 0-back condition. We found similar results in Kong's atlas with 200 ROIs. There is more dense connectivity in the frontoparietal cortex under 2-back conditions compared to 0-back and fixation conditions, resulting in larger purple modules (module k=8). There is a smaller sensory motor and attention-related module (blue module k=7) under 2-back compared with 0-back. 2-back condition also shows a suppressed default mode network (brown module k=5) compared with 0-back.

\subsubsection{Between subjects, subject-specific to group, and between conditions validations}
\begin{figure}[!ht]
\centering
\includegraphics[width=1\linewidth]{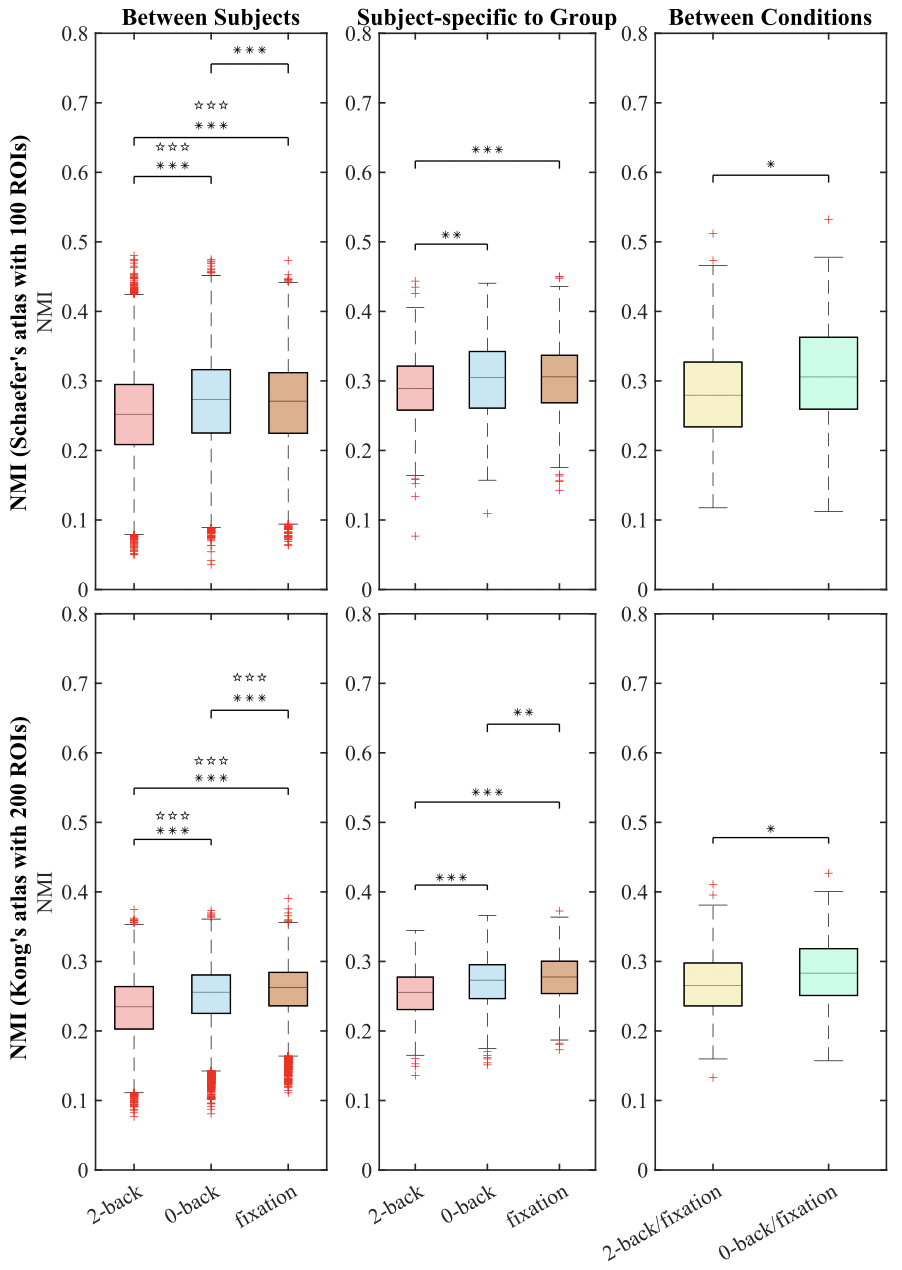}
\caption{\footnotesize Statistical analyses of NMI between subjects at individual level, NMI of subject-specific community memberships to the group average, and NMI across conditions based on Schaefer's atlas with 100 ROIs and Kong's atlas with 200 ROIs. The star denotes the significance level of the F-test for variance homogeneity (\ding{73}: $0.01\leq p<0.05$, \ding{73}\ding{73}: $0.001\leq p<0.01$, and \ding{73}\ding{73}\ding{73}: $p<0.001$). If the p-value of F-test is smaller than 0.05, Welch's t-test is used, otherwise, conventional two sample t-test is used. The asterisk indicates the statistical significance of t-test ($\ast$: $0.01\leq p<0.05$, $\ast$ $\ast$: $0.001\leq p<0.01$, and $\ast$ $\ast$ $\ast$: $p<0.001$). The p-values were corrected for multiple comparisons using the Benjamini-Hochberg false discovery rate (FDR) procedure.}
\label{2b0bfix}
\end{figure}
To evaluate the utility and stability of our LBM based method, we evaluated the community detection results between subjects under three task conditions using both Schaefer's atlas and Kong's atlas. We found that the estimated community structures between subjects are significantly more discrepant in terms of NMI of the community memberships between subjects under 2-back condition compared to that under 0-back condition as shown in Fig.\ref{2b0bfix}. We also evaluate similarity of subject-specific community structures to group-level community structure by NMI using these two atlas. Our result shows larger subject-specific to group difference of the community structures under 2-back condition. It implies that human brain with high load of working memory task results in larger inter-individual deviation of community structures. Finally, we evaluate the difference in the community structure between conditions. We found that the difference in the structure of the community is significantly greater between the 2-back and fixation than between the 0-back and fixation.

\section{Discussion}

The FC of human brain may show prominently discrepancies between different subjects \cite{Betzel2019, Bian2025} which are not only caused by changes in latent cognition including mental processes \cite{Taghia2018a,Hutchison2013} (e.g., thoughts, ideas, awareness, arousal, and vigilance) occurring at an unpredictable timescale during the resting state \cite{Friston2014, Razi2015, Razi2016, Razi2017, Friston2020} and brain activity responding to an external stimulus during task \cite{Cribben2012, Gonzalez-Castillo2018, Vidaurre2018}, but also due to non-neural physiological factors such as head motion, cardiovascular, and respiratory effects or the noise coming from the hardware instability \cite{Hutchison2013, Lurie2020}. The noise may affect the measurement of transient functional interaction between pairs of nodes, which will further result in significant changes in the community structure of brain networks. One outstanding problem is the unreliability of single-subject estimate of FC because it ignores information that is shared across individuals \cite{Lehmann2021}. Even in task fMRI, although the performance of participants is constrained by an external stimulus, the noise inevitably affects the metrics of individual FC. In this paper, we consider the group-level community structure to depict the hierarchical brain networks and quantify the inter-individual variability of FC during a task, which helps to alleviate the random influence of external non-neural noise.

The samples of community memberships generated by the MCMC sampler with Gibbs and M3 moves \cite{Nobile2007,Wyse2012} typically got stuck in different local modes for different runs of MCMC simulation, which implies biased estimation, due to only a single observation (the group-averaged connectivity) being taken into account. Modelling the individual FC rather than the group-averaged FC provides insight into the variability of both the observations themselves, and the variability in the community structure between subjects, and can also alleviate the problem of the sampler getting stuck in a local mode. However, in the hierarchical Bayesian framework, the variability of the estimated latent labels at individual-level modelling results in variation in the sizes of the communities, which in turn results in variation in the sizes of the blocks in LBM at individual level. Therefore, we do not infer the block parameters $\boldsymbol{\pi}$ in the group-level functional brain networks in this paper.

For the latent block model, other mixture models should also be reasonable for FC depending on the assumption of block parameters $\pi_{kl}$ in block $kl$. For example, we can also use Wishart distribution for the weighted FC or Bernoulli model for binarized connections with a threshold.

In this paper, we relax the assumption of fixing the number of communities $K$ like in spectral clustering, and assume it as a random variable that follows a Poisson distribution. We estimate the value of $K$, considering it a free parameter, via Bayesian inversion rather than using a model fitting strategy \cite{Bian2021}. The absorption and ejection moves integrated into the Metroplis-Hastings sampler enables variation of $K$ in the constructed Markov chain, so that the samples of both $K$ and $\boldsymbol{z}$ can be drawn from the collapsed posterior density at the individual level. In this case, the inference of latent label vector $\boldsymbol{z}$ is not constrained by the pre-determined number of communities, which makes the estimates of labels at individual level more flexible compared to the method using a fixed value of $K$. In addition, in contrast to (multilayer) modularity techniques, our methods deliver better community detection precision. Rather than using the modularity resolution parameter to adjust community number or size, the LBM framework achieves this by setting the maximum allowable number of communities.

Next, we discuss the group-level modelling of FC. Constructing group representative network by estimating a mean (group-averaged) FC \cite{Bian2021} ignores the variation of brain networks across individuals. Setting a threshold for FC \cite{Bian} may lose some important topological information about the networks. In this paper, the method based on hierarchical Bayesian modelling is able to characterize both averaged strength and inter-individual variation of weighted FC between brain regions across population, which retains the complete information of both individual and group-level FC.

In this work, we proposed a community structure-based multivariate Gaussian generative model to simulate synthetic data with different levels of inter-individual variation regarding to the true subject-specific latent labels and the inter-  and intra-connectivity densities for validation of community detection algorithms. The proposed generative model is explainable and can potentially be applied to validate a wide range of models and fMRI applications.

The SNR for fMRI data is reported by using many different definitions \cite{Welvaert2013}. Most real fMRI data studies use tSNR (temporal SNR) SNR=$\overline{S}/\sigma_{N}$ where $\overline{S}$ is the mean signal over time and $\sigma_{N}$ is the standard deviation of the noise. For example, the reported SNR values of the fMRI data in \cite{Welvaert2013} ranged from 0.35 to 203.6, that is, -4.5593dB to 23.0878dB. Many simulation studies use the standard deviation ratio between the activation signal and noise (e.g., SNR$=\sigma_{S}/\sigma_{N}$ or SNR $=10log_{10}(\sigma_{S}/\sigma_{N}$) in dB format), and the variance ratio (e.g., SNR$=\sigma_{S}^2/\sigma_{N}^2$ or SNR$=10log_{10}(\sigma_{S}^2/\sigma_{N}^2)$ in dB format). According to the simulation reference table in \cite{Welvaert2013}, the SNR of the block-designed experiment ranges from -27.02dB to 12.98 dB for changes in activation signal $1\%$, -21.00dB to 19.00dB for changes in activation signal $2\%$ and -13.04dB to 26.96dB for changes in activation signal $5\%$.

According to the tSNR map (Figure S8 and S9 in \cite{Barch2013}) of the fMRI of the work memory task used in our study, the tSNR ranges from 30 to 120, that is 14.7712dB to 20.7918dB, which is relatively high quality compared to other conventional fMRI scans. Three SNR values of 10dB, 5dB, and 0dB are enough for the current fMRI study of the working memory task. However, to generalize to broader real fMRI data, we added new experiments with more SNR values of -5dB, -10dB, -15dB and -20dB and to test the performance of our community detection methods. We find that the community detection performance lost accuracy in -15 dB and -20 dB, so levels of SNR lower than -20dB were not necessary to be tested.

The method proposed in this paper solves many of the problems in the field of community detection. However, there are still some limitations of the current work. Community detection using LBM is more accurate compared to conventional (multilayer) modularity models, but with relatively high computational complexity. This computational cost is mainly due to two aspects. The first one is the requirement of time steps for the convergence of Markov chain and drawing enough samples from the chain. The second is the computational cost of the label switching algorithm. In addition, we do not consider the overlapping communities within the scope of this paper, as it is a complicated question to evaluate the inter-individual variability about the community structure for overlapping communities. In our framework, community memberships are first estimated for individuals and then for the group representative networks. Given the noise levels in the fMRI data, subject-specific communities are likely to be driven by a combination of true inter-individual variability and noise if not constrained by group priors. To avoid this issue and minimise the noise levels, individual-level modelling techniques for fMRI that either use group priors or iterate between subject- and group-level estimations until a convergence point is reached are worth exploring in future studies.

The community detection works well in synthetic data where the connectivity is formed explicitly based on predefined known community structure. However, the communities are unknown in real functional brain networks. There is no standard algorithm for general community detection because the network architectures in real world are assumed to be generated from different latent generation processes. Community detection using LBM is based on the information of adjacency matrix which only measures the FC between pairs of nodes, but does not utilize the information of metadata or features on the nodes. In the future, it is worth exploring combining node metadata and FC for community detection.

%

\end{document}